%% file: EGmain.tex
\documentclass{egpubl}
\usepackage{eg2026}
 
\ConferencePaper        
\CGFccby

\ifpdf \usepackage[pdftex]{graphicx} \pdfcompresslevel=9
\else \usepackage[dvips]{graphicx} \fi
\usepackage{booktabs}
\usepackage[T1]{fontenc}
\usepackage{dfadobe}  
\usepackage{import}   
\usepackage{subfigure}  
\usepackage{pgffor}      
\usepackage{catchfile}   
\usepackage{subcaption} 
\usepackage{tikz}
\usepackage{pgfplots}
\usepackage{overpic}

\usepackage{caption}
\usepackage[ruled]{algorithm2e} 
\usepackage{amsmath}
\biberVersion
\BibtexOrBiblatex
\usepackage[backend=biber,bibstyle=EG,citestyle=alphabetic,backref=true]{biblatex} 
\electronicVersion
\PrintedOrElectronic
\newcommand{\rev}[1]{#1}
\usepackage{xcolor}
\definecolor{darkyellow}{RGB}{204, 153, 0}
\newcommand{\revtwo}[1]{#1}
\ifpdf \usepackage[pdftex]{graphicx} \pdfcompresslevel=9
\else \usepackage[dvips]{graphicx} \fi

\usepackage{egweblnk} 
\makeatletter
\copyrightTextTitPag{}
\copyrightTextRunPag{}
\def\ps@titlepage{%
  \let\@mkboth\@gobbletwo
  \def\@oddhead{}\def\@evenhead{}%
  \def\@oddfoot{}\def\@evenfoot{}%
}
\def\ps@headings{%
  \let\@mkboth\@gobbletwo
  \def\@oddhead{}\def\@evenhead{}%
  \def\@oddfoot{}\def\@evenfoot{}%
}
\makeatother

\title[\rev{PBR-Inspired} Controllable Diffusion for Image Generation]%
      {\rev{PBR-Inspired} Controllable Diffusion for Image Generation}

\author[Bowen Xue \& Giuseppe Claudio Guarnera \& Shuang Zhao \& Zahra Montazeri]
{\parbox{\textwidth}{\centering Bowen Xue$^{1}$\orcid{0000-0002-6628-577X} \quad Giuseppe Claudio Guarnera$^{2}$\orcid{0000-0002-7703-5194} \quad Shuang Zhao$^{3}$\orcid{0000-0003-4759-0514} \quad Zahra Montazeri$^{1}$\orcid{0000-0003-0398-3105} 
        }
        \\
{\parbox{\textwidth}{\centering 
$^1$University of Manchester, UK \quad $^2$University of York, UK \quad $^3$University of Illinois Urbana-Champaign, USA}
}
}

\begin{document}

\teaser{
 \includegraphics[width=1\linewidth]{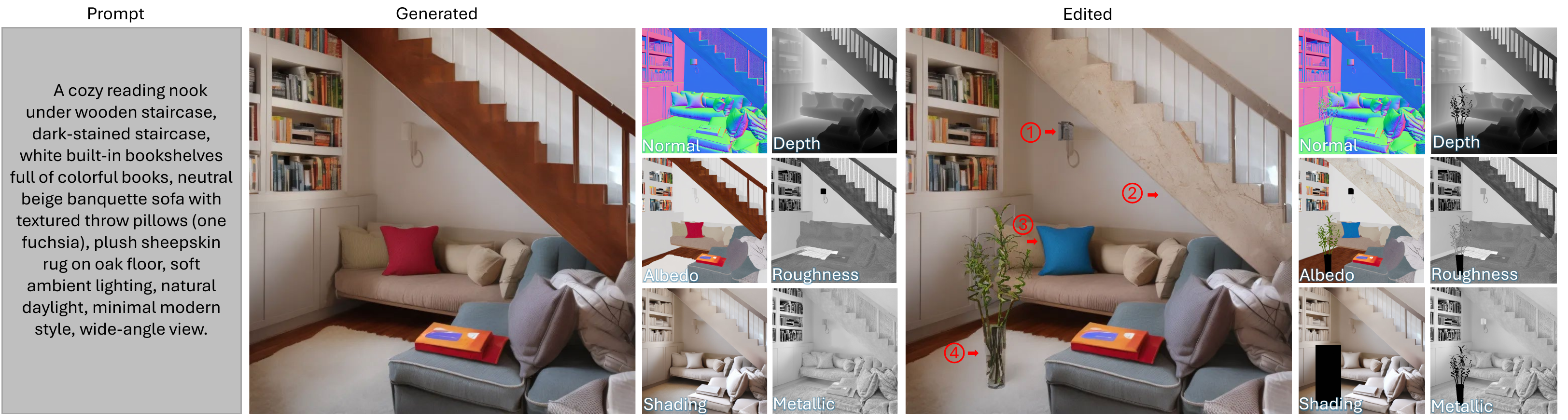}
 \centering
 \captionsetup{labelfont=bf, textfont=it}
  \caption{We propose a controllable text-to-image pipeline that begins by generating a \emph{G-buffer} (geometry buffer) for any given text prompt. The G-buffer encodes per-pixel geometry and material channels such as albedo, surface normals, depth, roughness, metallic, and shading. The left shows the input text prompt, the middle shows our generated result, and the right demonstrates channel-wise control to the predicted G-buffer. Users can control specific channels to achieve arbitrary edits such as surface property adjustments (enhancing the metallic quality of the wall lamp \textcircled{\raisebox{-0.2ex}{1}}), material replacements (wooden staircase to marble \textcircled{\raisebox{-0.2ex}{2}}), color modifications (changing the sofa cushion color from red to blue \textcircled{\raisebox{-0.2ex}{3}}), and object insertion (adding a \rev{vase} on the carpet \textcircled{\raisebox{-0.2ex}{4}}).}
\label{fig:teaser}
}

\maketitle
\begin{abstract}
   Despite recent advances in text-to-image generation, controlling geometric layout and \revtwo{PBR material properties} in synthesized scenes remains challenging. We present a pipeline that first produces a G-buffer (albedo, normals, depth, roughness, shading, and metallic) from a text prompt and then renders a final image through a \rev{PBR-inspired branch network}. This intermediate representation enables fine-grained control: users can copy and paste within specific G-buffer channels to insert or reposition objects, or apply masks to the irradiance channel to adjust lighting locally. As a result, real objects can be seamlessly integrated into virtual scenes. By separating user-friendly scene description from image rendering, our method offers a practical balance between detailed post-generation control and efficient text-driven synthesis. We demonstrate its effectiveness through quantitative evaluations and a user study with 156 participants, showing consistent human preference over strong baselines and confirming that G-buffer control extends the flexibility of text-guided image generation.

\ccsdesc[500]{Computing methodologies~Reflectance modeling}
\ccsdesc[300]{Computing methodologies~Image-based rendering}
\printccsdesc   
\end{abstract}  
\input{Main}
\end{document}

%% file: Main.tex
\section{Introduction}
The ability to generate and edit photorealistic scenes with precise control over \revtwo{PBR} material properties remains a challenge in computer graphics. While recent text-to-image diffusion models~\cite{Ramesh21, Saharia22, Rombach22} excel at generating visually stunning images, they operate as end-to-end mappings that obscure the underlying scene structure. This black-box nature fundamentally limits their utility in professional workflows where artists need fine-grained control over specific scene elements—adjusting material properties, modifying lighting, or editing geometry—without regenerating the entire image. For a visual preview of our controllable pipeline and edits, see Figure~\ref{fig:teaser}.

This limitation stems from a structural mismatch: the direct text-to-RGB mapping entangles geometry, materials, and lighting, making precise, channel-wise control impractical. Current diffusion models learn a direct mapping from text to RGB pixels, entangling geometry, materials, and lighting into a single, inseparable representation. When users attempt to modify prompts for targeted edits (e.g., ``make the table metallic''), the model cannot disentangle which pixels correspond to the table's material properties versus its geometry or the surrounding lighting. Recent efforts to enhance controllability have taken two divergent paths, each with fundamental limitations. Structural control methods (ControlNet~\cite{zhang2023adding}, T2I-Adapter~\cite{10.1609/aaai.v38i5.28226}) successfully inject spatial conditions but operate at the image level, missing the rich semantic structure of scene properties. Conversely, neural-based decomposition methods (\rev{\revtwo{RGB$\leftrightarrow$X}}~\cite{zeng2024rgb}, neural rendering approaches~\cite{liang2025diffusionrenderer}) attempt to recover scene structure but \rev{specialize the diffusion prior to the indoor domain via fine-tuning on relatively small indoor datasets. \revtwo{RGB$\leftrightarrow$X} also supports editing and relighting capabilities}. This division suggests a critical gap: can we design an architecture that combines the generative power of pretrained diffusion models with explicit \revtwo{PBR} scene decomposition that allows detailed modification?

Our key insight is that this integration becomes feasible by carefully considering where and how to inject structure into the diffusion process. We identify two critical design decisions. First, we recognize that the latent space of diffusion models offers unique advantages for structured generation. Unlike image-space methods that must encode and decode through information-losing transformations, latent-space operation preserves the rich distributional structure learned during pretraining. This observation motivates our Latent ControlNet—a modified architecture that directly processes latent representations to generate G-buffers without intermediate image decoding. This enables the generation of consistent multi-channel outputs (albedo, normal, depth, roughness, metallic, shading). Second, we observe that the rendering equation naturally suggests a branch architecture for the inverse problem. Rather than learning a plain end-to-end mapping from G-buffers to images, we design a network that mirrors the \revtwo{PBR} image formation process: separate pathways for geometry (normals, depth), materials (albedo, roughness, metallic), and lighting (shading). This factorization provides an inductive bias that improves learning efficiency—our experiments reveal a 76\% reduction in reconstruction error compared to end-to-end alternatives.

These insights lead to a two-stage framework that bridges generative modeling and \rev{PBR-inspired} rendering. We separate scene structuring from image rendering: a Latent ControlNet first produces a multi-channel G-buffer, and a \rev{PBR-inspired} branch renderer then maps it to RGB, enabling predictable, channel-wise control. Stage one generates complete G-buffers from text prompts using our Latent ControlNet, while stage two renders these buffers into images through our branch network. Crucially, this decomposition enables intuitive post-generation editing: users can directly manipulate individual channels with predictable, localized effects on the final image.

In summary, our contributions include:\\
(1) Latent ControlNet Architecture: A Latent ControlNet that operates directly in diffusion latent space, enabling multi-channel G-buffer generation while preserving information fidelity. (Sec.~\ref{sec:texttog})\\
(2) \rev{PBR-Inspired Branch Renderer}: A \rev{branch network} that mirrors the image formation process through separate geometry, material, and lighting pathways, achieving superior reconstruction.(Sec.~\ref{sec:gbufferrenderingnetwork})\\
(3) Practical Editing Framework: A system to enable intuitive, \rev{PBR-inspired} editing of diffusion-generated images through G-buffer manipulation, validated through extensive user studies.

\section{Related Work}
\textbf{Text-to-Image Generation.}
Text-to-image generation has advanced considerably through GANs and diffusion models. GAN-based methods demonstrated the feasibility of synthesizing images from textual descriptions but often suffered from low resolution and limited semantic alignment~\cite{Reed16, Zhang17}. Attention-based architectures improved the correlation between text embeddings and visual content~\cite{Xu18}. In parallel, diffusion models offered better coverage of the data manifold and reduced mode collapse~\cite{Dhariwal21}, leading to large-scale systems like DALL·E~\cite{Ramesh21} and GLIDE~\cite{Nichol22}, which introduced classifier-free guidance. Latent diffusion approaches enabled high-resolution outputs at reduced cost~\cite{Rombach22}, forming the basis of Stable Diffusion, while Imagen further improved photorealistic generation through cascaded diffusion~\cite{Saharia22}. Despite these advances, post-generation editing of specific scene elements remains challenging.

\textbf{Screen-Space Rendering.}
\rev{The G-buffer concept, originally introduced by Saito and Takahashi~\cite{saito1990comprehensible} for comprehensible rendering of 3D shapes, has become fundamental to deferred shading pipelines.} Screen-space rendering bridges purely 2D methods and full 3D reconstructions, capturing partial spatial data via layered depth images or G-buffers~\cite{Buehler01, Penner17}. This enables moderate viewpoint changes and scene manipulation without the complexity of volumetric geometry. For instance, \cite{Hedman18} supported free-viewpoint navigation with multi-view blending.

\textbf{Neural Rendering.}
Neural rendering synthesizes novel views or edited scenes from volumetric or surface-based data, bridging computer graphics and machine learning. Hierarchical neural materials improve multi-scale SVBRDF fidelity~\cite{Xue2024NeuralMaterials}. NeRFs~\cite{Mildenhall20} first demonstrated high-fidelity view synthesis later extended to dynamic scenes~\cite{Park21}, and anti-aliased systems for unbounded environments~\cite{Barron22}. Surface-based methods~\cite{Lombardi19, Tretschk21} utilize explicit geometry or point-based representations for realistic rendering but often demand substantial data and computation, complicating fine-grained edits. DiffusionRenderer~\cite{liang2025diffusionrenderer} proposes a combined framework for both neural inverse and forward rendering using video diffusion models. 

\textbf{Neural Image Synthesis from G-buffer.}
\rev{This line of work builds on foundational intrinsic image decomposition~\cite{barrow1978recovering, bell2014intrinsic} and \revtwo{PBR}-based shading models~\cite{burley2012physically}.} Several approaches learn image synthesis from G-buffers or intermediate decompositions. Deep Shading~\cite{Nalbach2017DeepShading} infers effects like ambient occlusion and subsurface scattering via CNNs. \cite{Zhu2022ScreenSpaceRayTracing} propose screen-space ray tracing from intrinsic channels, and \revtwo{RGB$\leftrightarrow$X}~\cite{zeng2024rgb} fine-tunes a Stable Diffusion model to render intermediate decompositions. 

\textbf{Editing and Relighting.}
Neural relighting can rely on explicit representations~\cite{Griffiths2022Relight,Pandey2021Relight,Yu2020Relight} or implicit ones~\cite{Rudnev2022RelightImplicit,Wang2023RelightImplicit}, typically limited to simpler lighting conditions. Recent works have explored diffusion models for lighting control and intrinsic decomposition. DiLightNet~\cite{Zeng_2024} provides fine-grained lighting control for diffusion-based image generation, while LightIt~\cite{kocsis2024lightit} uses light maps to control image generation. "Intrinsic Image Diffusion"~\cite{kocsis2024iid} and IntrinsicDiffusion~\cite{IntrinsicDiffusionluo} both employ diffusion models for intrinsic image decomposition. Diffusion-handles~\cite{pandey2024diffusion} applies a diffusion-based model for object manipulation. In recent years, a variety of extensions have been proposed to improve controllability of diffusion model, yet previous works either inject structural conditions (ControlNet \cite{zhang2023adding}, T2I-Adapter \cite{10.1609/aaai.v38i5.28226}, GLIGEN \cite{10203593}) but lack fine-grained material and lighting control, or enable local editing (Prompt-to-Prompt \cite{hertz2022prompt}, DiffEdit \cite{couairon2022diffeditdiffusionbasedsemanticimage}, ZONE \cite{10658119}) and concept insertion (DreamBooth \cite{ruiz2022dreambooth}, AnyDoor \cite{chen2023anydoor}) but remain limited to object-level modifications without true channel-wise refinement capabilities. DiffusionRenderer \cite{liang2025diffusionrenderer} leverage diffusion models for neural rendering which offers unified inverse–forward framework for temporally consistent video decomposition and re-rendering of existing content; however our work embeds rendering knowledge into generative models for text-driven control over \revtwo{PBR} properties to create controllable content from scratch.
\rev{IntrinsicEdit~\cite{10.1145/3731173} proposes a training-free optimization framework built on top of pre-trained \revtwo{RGB$\leftrightarrow$X} models. While IntrinsicEdit focuses on editing existing images, our work trains a dedicated text$\rightarrow$G-buffer$\rightarrow$RGB backbone for generating new content. These approaches are orthogonal: IntrinsicEdit could be applied on top of our backbone as a complementary editing layer. Recent general-purpose editing models such as Flux \cite{labs2025flux1kontextflowmatching} and Qwen-Edit \cite{wu2025qwenimagetechnicalreport} achieve impressive results but operate directly on RGB without exposing intermediate scene representations, targeting a different use case from our channel-wise controllable generation.}

\begin{figure*}[htbp]
  \centering
  \includegraphics[width=1\linewidth]{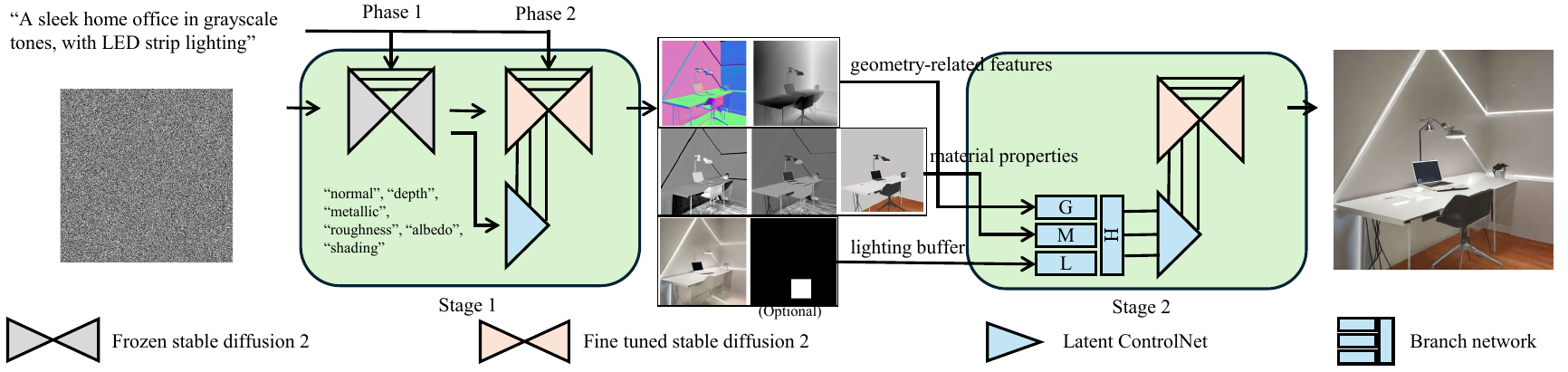}
  \caption{Overview. Our pipeline begins with a random noise sample and a text prompt. These inputs are processed by the stage-1 network, which consists of two denoising steps: first, a frozen Stable Diffusion 2 model (in gray), second, a fine-tuned Stable Diffusion 2 model augmented with ControlNet. Stage 1 produces a G-buffer comprising albedo, normal, depth, shading, roughness, and metallic. These channels are then grouped and passed to the stage-2 network, where an optional mask is used for object movement or insertion. Each group is processed by specialized sub-networks, fused by a final grouping module, and then fed into another ControlNet-equipped, fine-tuned Stable Diffusion 2 model to generate the final RGB output.}
  \label{fig:network}
\end{figure*}

\section{Text-to-G-Buffer Generation via Latent-Space ControlNet}
\label{sec:texttog}
In this section, we present our two-phase, diffusion-based network integrating a latent-space ControlNet for direct G-buffer generation from text prompts. The entire pipeline is depicted in Figure~\ref{fig:network}.

Our dataset is several orders of magnitude smaller than the original Stable Diffusion training dataset and focuses predominantly on indoor scenes. Fine-tuning the diffusion model in such a narrow subset can result in overfitting and catastrophic forgetting, as shown in Figure~\ref{fig:comparisoncontrolnet}. 

To generate G-buffers without overwriting the pretrained generative mapping, we adopt a two-phase network design similar to \cite{xue2024reflectancefusion}. In phase 1, we use a frozen diffusion model to preserve the pretrained generative capabilities. In phase 2, we introduce a latent-space ControlNet structure rather than retraining the diffusion model itself. Toward the last 5 epochs of training, we unfreeze the main diffusion model, further improving G-buffer results. 

In phase 2, directly fine-tuning the stable diffusion on a small dataset \(D\) needs to relearn the entire G-buffer mapping by solving
\begin{equation}
  \theta^* 
  = \arg\min_{\theta}
    \frac{1}{|D|}\sum_{(x,y)\in D}
      \ell\bigl(F(x;\theta),\,y\bigr),
\end{equation}
where all UNet parameters \(\theta\) are being updated. In contrast, we retain the pretrained UNet backbone \(\mathcal F^*(x)=F(x;\theta^0)\) (with frozen weights \(\theta^0\)) and learn only a lightweight ControlNet \(g_\phi\) via
\begin{equation}
  \phi^* 
  = \arg\min_{\phi}
    \frac{1}{|D|}\sum_{(x,y)\in D}
      \ell\!\bigl(g_\phi(\mathcal{F}^*(x)),\,y\bigr).
\end{equation}
Here \(\phi\) denotes parameters of \(g_\phi\). This formulation preserves the backbone's original generative diversity, yielding better sample efficiency and reduced overfitting compared to full fine-tuning.

By preserving the pretrained parameters and introducing a relatively small ControlNet, we retain the latent function and confine new adaptations to a lower-dimensional space. This strategy mitigates catastrophic forgetting and improves generalization on small-data tasks compared to fully tuning all parameters. As shown in Figure~\ref{fig:comparisoncontrolnet}. Let $\mathcal{E}$ denote the control net encoder, $\mathcal{D}$ the diffusion model VAE decoder. The standard ControlNet pipeline is:
\begin{equation}
  z_c \xrightarrow{\mathcal{D}} x_c \xrightarrow{\mathcal{E}} z_c' \xrightarrow{g} \hat{f}_c 
\end{equation}
where $z_c$ is the latent representation control signal, $x_c$ is the control image, $g$ is the standard ControlNet. Our architecture simplifies to:
\begin{equation}
\label{ourcontrol}
  z_c \xrightarrow{g_\phi} \hat{f}_c 
\end{equation}
By operating directly in the latent space, we avoid \rev{an additional decode--encode step that may discard information. As qualitative motivation, the data processing inequality~\cite{cover1999elements} for the Markov chain $z_c \to x_c \to z_c'$ gives:}
\begin{equation}
I(z_c; Y) \geq I(z_c'; Y)
\end{equation}
where $Y$ represents the target G-buffer and $I(\cdot;\cdot)$ denotes mutual information. \rev{This suggests that the additional decode-encode cycle cannot increase information about the target.}

Furthermore, the decoder-encoder cycle $\mathcal{D} \to \mathcal{E}$ may discard latent information that is not explicitly represented in RGB space but is crucial for G-buffer generation. For instance, the latent code $z_c$ may contain implicit geometric or material properties that are lost when projected to RGB and cannot be fully recovered by $\mathcal{E}$.

Our Latent ControlNet, operating directly on $z_c$, \rev{avoids an additional decode–encode step that can discard information. Empirical validation of this design is provided in Figures~\ref{fig:comparisoncontrolnet} and~\ref{fig:comparisonlatentcontrolnet}.}
This design is particularly advantageous for G-buffer generation, where we need to extract structured scene information (geometry, materials, lighting) that may be implicitly encoded in the latent space but not fully visible in the RGB representation. Our method also reduces computational complexity by eliminating the control signal decoding and encoding phase and keeps representation consistency. When the control signal resides in the same latent space as the diffusion model, the conditional generation process becomes more direct.

\input{ablationcontrolnet}

\section{G-buffer Rendering Network}
\label{sec:gbufferrenderingnetwork}
Our G-buffer to image rendering pipeline leverages a fine-tuned Stable Diffusion model with a ControlNet backbone, which natively supports three-channel conditional inputs. However, our ControlNet input consists of 13 stacked channels: albedo (3), normal (3), roughness (1), shading (3), metallic (1), depth (1), and a mask (1) that indicates regions where new objects are inserted or existing objects are moved. Simply concatenating these channels tends to cause poor performance and training instability; moreover, our extra channels form a complex, multi-component G-buffer rather than additional color images. Hence, we prepend a multi-layer CNN module to the input of ControlNet to extract low-level features from this multi-channel input. This strategy enhances compatibility with the original architecture and stabilizes training.

\rev{\textbf{Single-Bounce Baseline.}} A simplified version of Kajiya's rendering equation~\cite{10.1145/15922.15902} for a surface point $p$ and outgoing direction $\omega_o$ (where $f_r$ is a local BRDF, $L_i$ is the incident radiance) can be written as:
\begin{equation}
\label{eq:kajiya_appendix}
L_o(p,\,\omega_o) = \int_{\Omega} f_r\bigl(p;\,\omega_i,\omega_o\bigr)\; L_i\bigl(p;\,\omega_i\bigr)\; (\mathbf{n}\cdot\omega_i) \;d\omega_i,
\end{equation}
For simplicity, we omit secondary bounces, subsurface scattering, and emissive effects.

\textbf{Microfacet BRDF Decomposition.} Under microfacet theory (Cook--Torrance~\cite{10.1145/357290.357293}), the BRDF $f_r$ factorizes into Fresnel $F$, normal distribution $D$, and a geometric masking term $G$:
\begin{equation}
\label{eq:microfacet_appendix}
f_r(\omega_i,\omega_o;\,\mathbf{x}_m,\mathbf{x}_g)
~=~
\frac{F(\mathbf{x}_m,\mathbf{h})
       \;\,D(\mathbf{x}_m,\mathbf{h})
       \;\,G(\mathbf{x}_g,\omega_i,\omega_o)
}{
   4\,(\mathbf{n}\cdot\omega_i)\,(\mathbf{n}\cdot\omega_o)
},
\end{equation}
where $\mathbf{x}_m$ (material) includes roughness, metallic, and $\mathbf{x}_g$ (geometry) includes the macroscopic normal $\mathbf{n}$. Meanwhile, $L_i(\omega_i)$ may be regarded as a function of $\mathbf{x}_l$ (e.g., environment maps, shadow buffers). Substituting Eq.~\eqref{eq:microfacet_appendix} into Eq.~\eqref{eq:kajiya_appendix} gives:
\begin{equation}
\label{eq:app_factorization_main}
L_o=\! \int_{\Omega} 
\frac{F(\mathbf{x}_m,\mathbf{h}) D(\mathbf{x}_m,\mathbf{h}) G(\mathbf{x}_g,\omega_i,\omega_o)}
{4(\mathbf{n}\!\cdot\!\omega_i)(\mathbf{n}\!\cdot\!\omega_o)}\;
L_i(\omega_i)\;(\mathbf{n}\!\cdot\!\omega_i)\, d\omega_i .
\end{equation}
Thus, geometry ($\mathbf{x}_g$), material ($\mathbf{x}_m$), and lighting ($\mathbf{x}_l$) appear in separate factors. This partial independence motivates a network design that processes these components in separate branches, often requiring fewer parameters and yielding more stable training.

\textbf{G-buffers and Branch Networks.}
In practice, we store $\{\mathbf{n}, \mathbf{d}\}$ in a \emph{geometry buffer}, $\{\mathbf{A}, r, m\}$ in a \emph{material buffer}, and $\mathbf{x}_{l}$ (shading, shading mask) in a \emph{lighting buffer}, where $\mathbf{n}$ denotes the normal, $\mathbf{d}$ the depth, $\mathbf{A}$ the albedo (base color), $r$ the roughness, and $m$ the metallic. The factorization in Eq.~\eqref{eq:app_factorization_main} provides theoretical justification for our branch network design. Instead of learning a monolithic function:
\[
F: \bigl(\mathbf{n}, \mathbf{d}, \mathbf{A}, r, m, \mathbf{x}_{l}\bigr)
~\mapsto~
\mathbf{L}_o,
\]
we factor it into three sub-networks:
\begin{equation}
\label{eq:factor_network_threebranch}
\mathbf{L}_o
~=~
\mathcal{H}\Bigl(
   \mathcal{G}\bigl(\mathbf{n},\,\mathbf{d}\bigr),
   \;\mathcal{M}\bigl(\mathbf{A},\,r,\,m\bigr),
   \;\mathcal{L}\bigl(\mathbf{x}_{l}\bigr)
\Bigr),
\end{equation}
where $\mathcal{G}(\cdot)$ extracts geometry-related features, $\mathcal{M}(\cdot)$ operates on material properties, $\mathcal{L}(\cdot)$ processes the lighting buffer, and $\mathcal{H}(\cdot)$ fuses these intermediate embeddings to predict the final $\mathbf{L}_o$.


\revtwo{\textbf{Branch Network Architecture.}
Each branch ($\mathcal{G}$, $\mathcal{M}$, $\mathcal{L}$) takes as input 
$512\times512$ resolution maps: geometry (normal + depth, 4 channels), 
material (albedo + roughness + metallic, 5 channels), and lighting 
(shading + mask, 4 channels), respectively. Each branch employs a 4-layer 
CNN encoder with output channels [64, 128, 256, 512] using $3\times3$ convolutions 
followed by GroupNorm(8) and SiLU activation, where layers 2 and 3 use 
stride-2 downsampling. At the bottleneck ($128\times128\times512$), features 
are patchified (patch size 2, yielding $64\times64$ tokens with 512 dimensions) 
and processed by 2 DiT-style Transformer blocks~\cite{Peebles2022DiT}; each block consists 
of a pre-norm multi-head self-attention layer (dim=512, heads=8) and an MLP 
(expansion ratio 4, GELU activation) with residual connections. The fusion 
module $\mathcal{H}$ concatenates the three branch features along the channel 
dimension ($64\times64$ tokens $\times$ 1536 dim), projects to 512 dimensions, 
processes through one additional DiT block, reshapes from $(4096, 512)$ tokens to $(512, 64, 64)$ spatial features, and applies a zero-initialized $1\times1$ convolution to 320 
channels before feeding into the ControlNet backbone. The branch network 
totals 30.8M parameters.}

This decomposition aligns with the principle that incorporating domain knowledge as structural priors improves learning efficiency~\cite{47094,Higgins2016betaVAELB}. \rev{By approximating the single-bounce microfacet model as a structural prior}, our network design not only reduces the parameter space but also provides an inductive bias that empirically improves training stability.

\input{comparediffusionhandles}

\begin{figure}[htbp]
    \centering
    \includegraphics[width=1\linewidth]{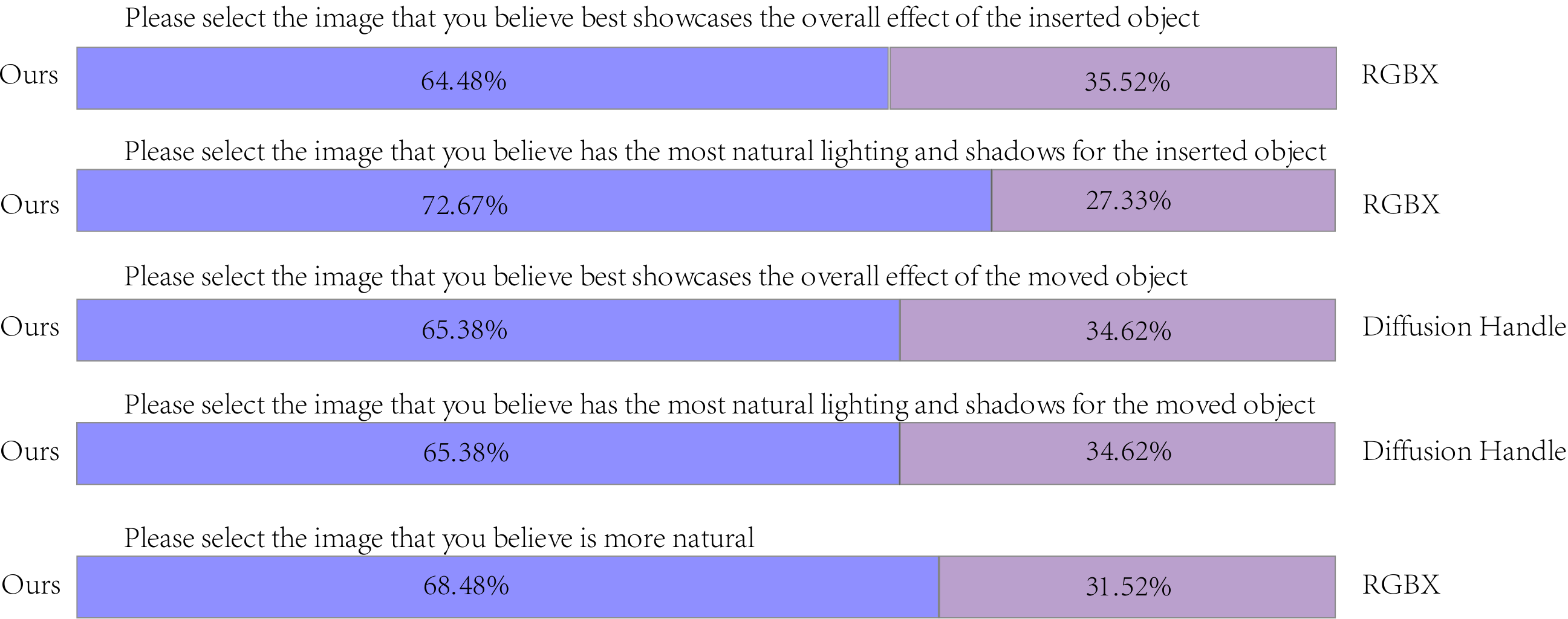}
    \caption{
        User study results (156 participants). Each bar illustrates the percentage of participants who preferred either our method or the baseline methods across various evaluation criteria. Participants compared pairs of static images generated by our method and the respective baseline (\revtwo{RGB$\leftrightarrow$X} or Diffusion Handles). Three evaluation questions were utilized for comparisons with \revtwo{RGB$\leftrightarrow$X}, while two questions were employed for comparisons with Diffusion Handles due to technical limitations in rendering buffers with Diffusion Handles.
    }
    \label{fig:user_study_bar}
\end{figure}

\input{imagecompare2.tex}

\section{Implementation Details}
\label{sec:implementation}

\textbf{Editing and Inpainting.}
When performing inpainting or moving objects, we directly copy the target object into the albedo, normal, roughness, depth, and metallic channels. For the shading map, we fill the edited region with black (i.e., zero) to indicate that these areas need to be recalculated, and simultaneously create a mask channel. In this mask, \texttt{mask}=1 denotes unmodified regions, while \texttt{mask}=0 indicates edited regions. The same procedure applies to object movement: we update the albedo and other channels, but for the shading channel, the network is guided by the mask to re-estimate lighting where needed.

\rev{\textbf{One-Click Object Insertion.}
For object insertion, users click on the target object region in the albedo map. We apply edge-based region selection to automatically identify the object boundaries. The selected region is then automatically propagated via copy-paste to all G-buffer channels (albedo, normal, roughness, depth, metallic). For the shading channel, we set the edited region to zero and correspondingly set mask=0 in these areas. Stage-2 then re-synthesizes the shading conditioned on the updated material channels, ensuring more consistent lighting for the inserted object. This workflow makes object insertion a mostly one-click operation.}

\textbf{Dataset.}
The original diffusion model was trained on an enormous dataset (2.47 billion samples), whereas our indoor-focused data is just $0.005\%$ of that size. To mitigate overfitting, we combine InteriorVerse\cite{zhu2022learning} (50k+ samples with albedo, normal, roughness, depth, and metallic) and Hypersim\cite{roberts:2021} (70k+ samples with shading/irradiance but lacking roughness and metallic), yielding over 120k samples in total.

\textbf{Training Procedure.}
Our implementation uses a ControlNet (378M parameters) paired with a UNet (865M parameters), with ControlNet representing approximately 44\% of UNet's size. We employ the mean squared error (MSE) as the loss function. All channels from the G-Buffer are normalized to the range $[0, 1]$. Specifically, the shading and target images are normalized based on the 99\% valid pixel range, while the depth channel undergoes logarithmic normalization. The network is trained for a total of 30 epochs. During the first 25 epochs, only the ControlNet is trained, with the main diffusion UNet kept \emph{frozen} to preserve its large-scale generative knowledge. In the final 5 epochs, we \emph{unfreeze} the main UNet. The initial learning rate is set to $5 \times 10^{-6}$ with a linear decay schedule, and the batch size is 16. The entire training process, executed on four A100 GPUs, requires approximately 150 hours. G-buffer generation model and rendering model trained separately.

\textbf{Mask-Guided Fine-Tuning.}
Note that our second-stage network (ControlNet + UNet) accepts albedo, normal, roughness, metallic, shading, and depth channels, plus an optional mask channel. In the first 10 epochs, the mask is set to 1 everywhere (no masking). Later, we gradually introduce regions where \texttt{mask}=0, forcing the shading there to be set to zero. This signals the model to re-estimate lighting in those areas, facilitating flexible edits in the final G-buffer. Once the network generates a complete G-buffer, each map (e.g.\ albedo, normal, roughness) can be freely edited, and for the shading channel specifically, we rely on the mask to indicate which parts need recomputation.
Our edited result can consider the shading effect outside the masked region. While the mask indicates regions needing shading recomputation, the network considers global illumination effects extending beyond the masked region. The mask serves as a guide for the network to focus on areas requiring direct shading updates, but the network architecture captures inter-reflections and indirect lighting through learned representations. The white box in figure is for visualization only—not a network input. In Figure~\ref{fig:comparisonmove}, the chair’s reflection extends beyond the masked region (last image), and the cushions cast reflections outside the mask boundaries (middle two examples).

\begin{figure*}[t]
    \centering
    \small
    \setlength{\tabcolsep}{1pt}
    \renewcommand{\arraystretch}{1}
    
    \begin{tabular}{@{}c@{\hspace{1pt}}c@{\hspace{1pt}}c@{\hspace{1pt}}c@{}}
        \multicolumn{2}{c}{\parbox{0.48\textwidth}{\centering\footnotesize  Prompt: Elegant bedroom with  \textbf{mint green} tufted headboard, \textbf{light blue floral wallpaper, pink bedding}, wooden nightstand, herringbone flooring, soft pastel color scheme, dreamy atmosphere, interior design photography}} & 
        \multicolumn{2}{c}{\parbox{0.48\textwidth}{\centering\footnotesize Prompt: A peaceful nursery retreat with walls \textbf{the color of morning mist, furniture as pure as fresh linen, and drapery that resembles vanilla cream}. The  \textbf{soft white wooly rug} contrasts beautifully with the rich chestnut floors}} \\[20pt]
        
        \begin{overpic}[width=0.249\textwidth]{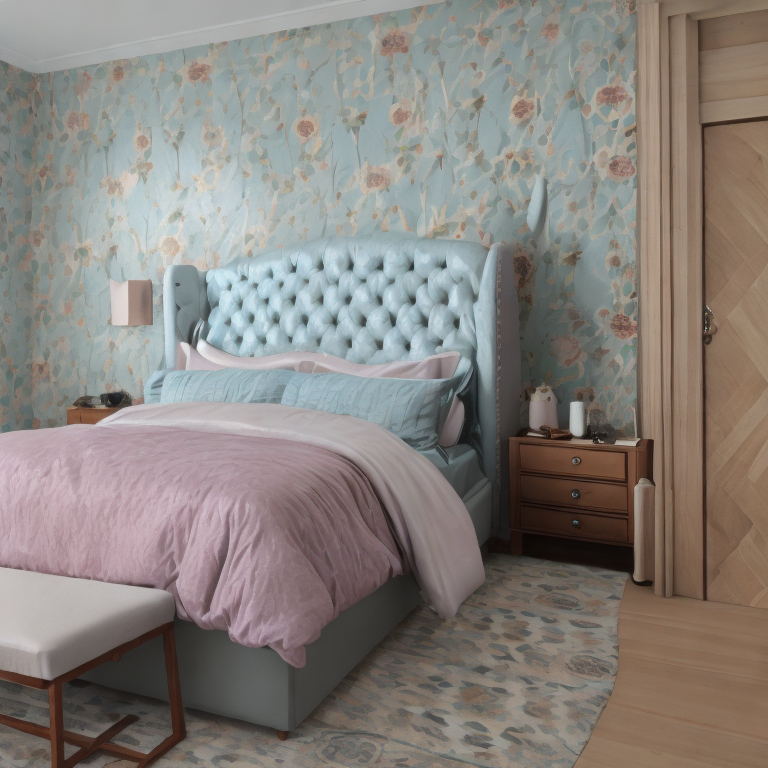}
            \put(0,67){\includegraphics[width=0.08\textwidth]{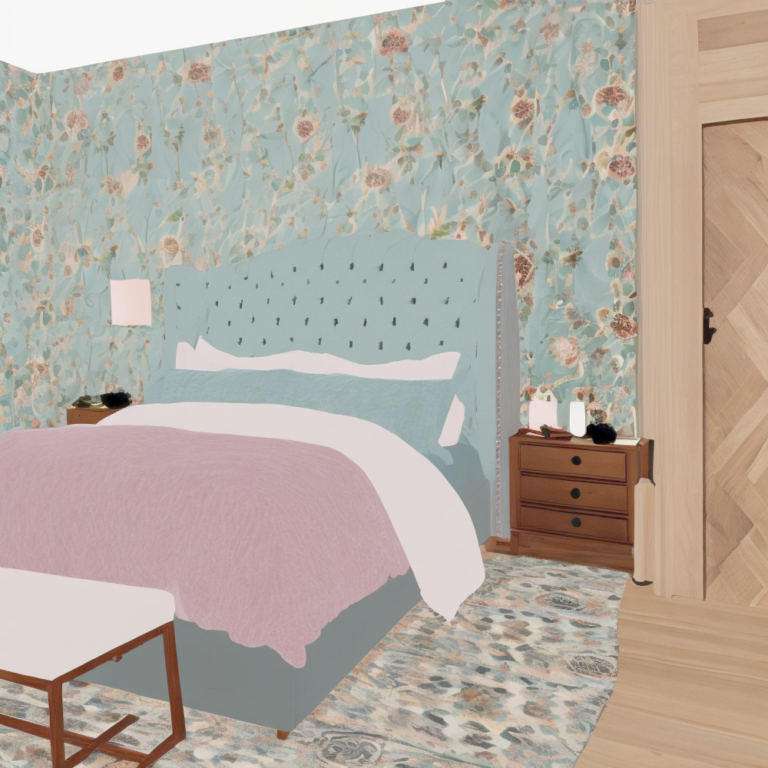}}
        \end{overpic} &
        
        \begin{overpic}[width=0.249\textwidth]{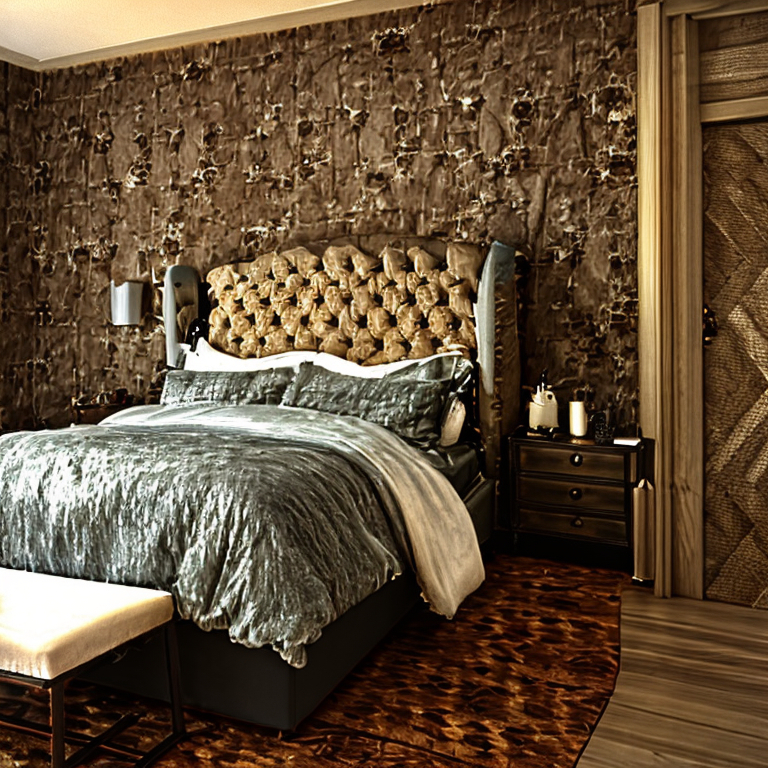}
            \put(0,67){\includegraphics[width=0.08\textwidth]{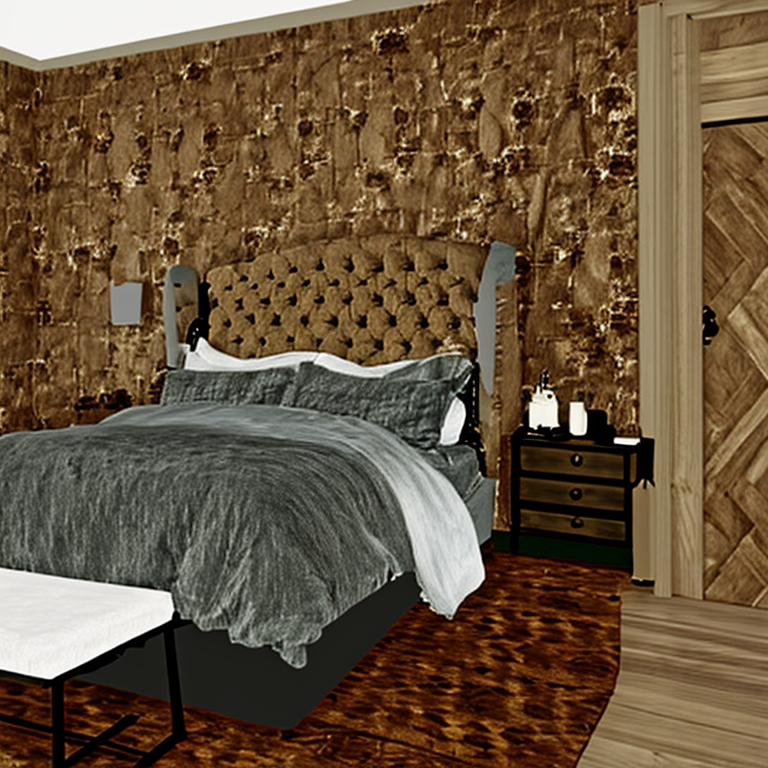}}
        \end{overpic} &
        
        \begin{overpic}[width=0.249\textwidth]{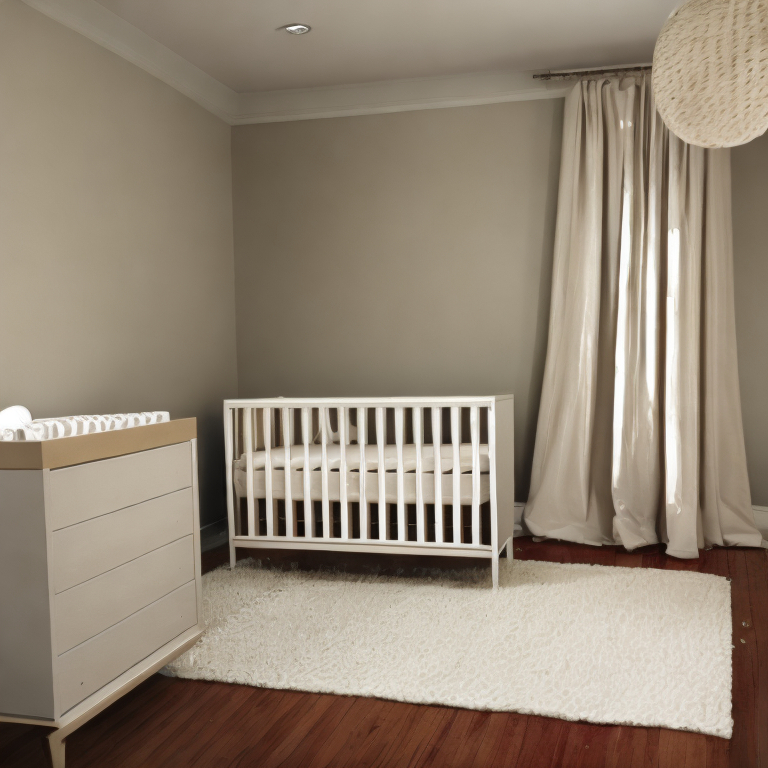}
            \put(0,67){\includegraphics[width=0.08\textwidth]{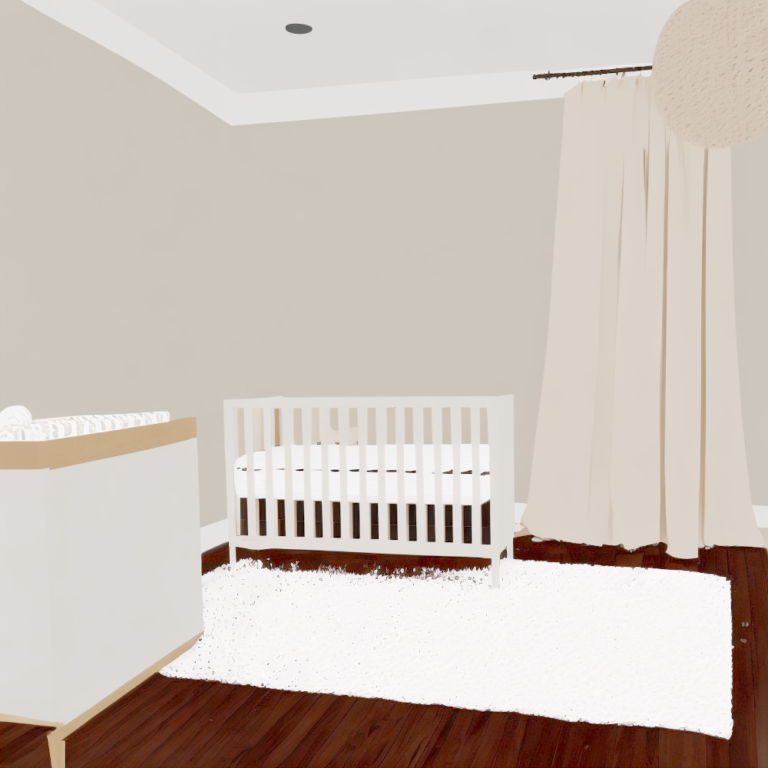}}
        \end{overpic} &
        
        \begin{overpic}[width=0.249\textwidth]{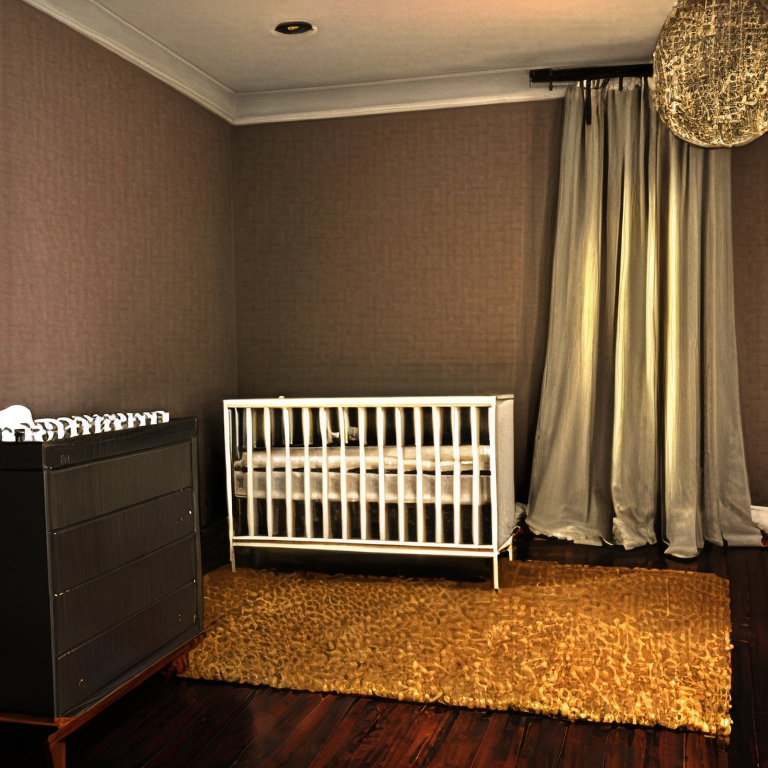}
            \put(0,67){\includegraphics[width=0.08\textwidth]{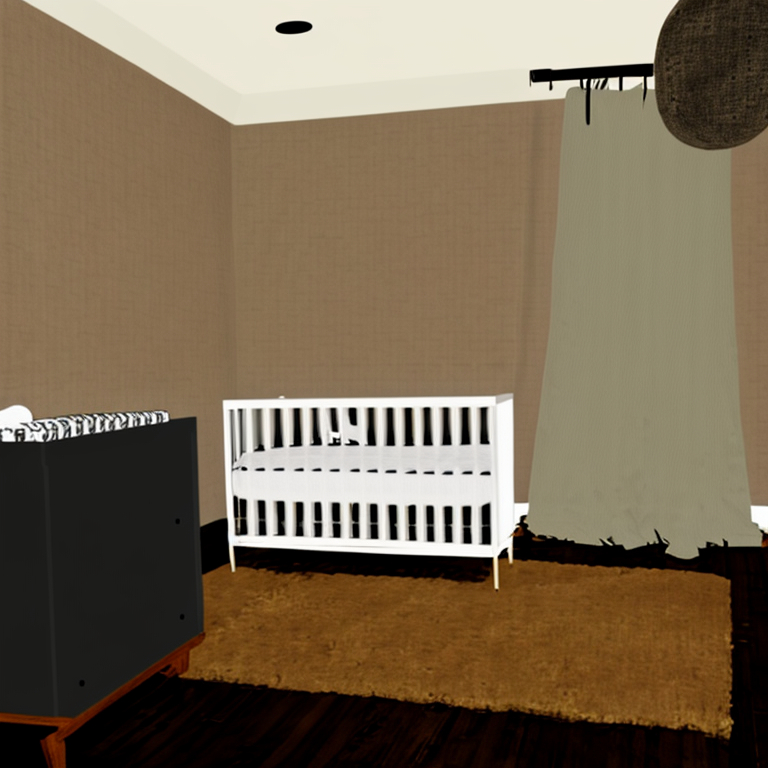}}
        \end{overpic} \\
        
        {\footnotesize Latent-space ControlNet} & {\footnotesize Standard Controlnet} & 
        {\footnotesize Latent-space ControlNet} & {\footnotesize Standard Controlnet}  \\
        \\
        
        \multicolumn{2}{c}{\parbox{0.48\textwidth}{\centering\footnotesize Prompt: A centuries-old European castle courtyard featuring  \textbf{a perfectly manicured formal garden}. Hedge mazes and topiary figures define winding gravel paths. In the background, stone towers and gothic spires add grandeur to the once-fortified domain.}} & 
        \multicolumn{2}{c}{\parbox{0.48\textwidth}{\centering\footnotesize Prompt: Cozy indoor plant corner with white walls. Multiple hanging plants in macramé holders. Monstera and trailing plants. \textbf{Small purple sofa with colorful cushions}. Terra cotta planters. Natural light from window. Minimalist room with jungle vibes.}} \\[20pt]

        \begin{overpic}[width=0.249\textwidth]{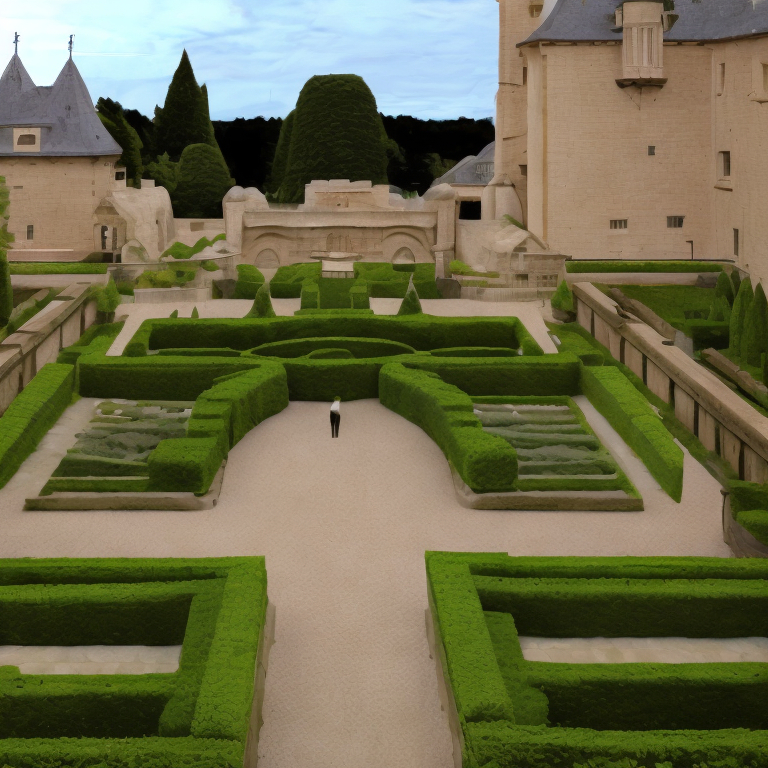}
            \put(0,67){\includegraphics[width=0.08\textwidth]{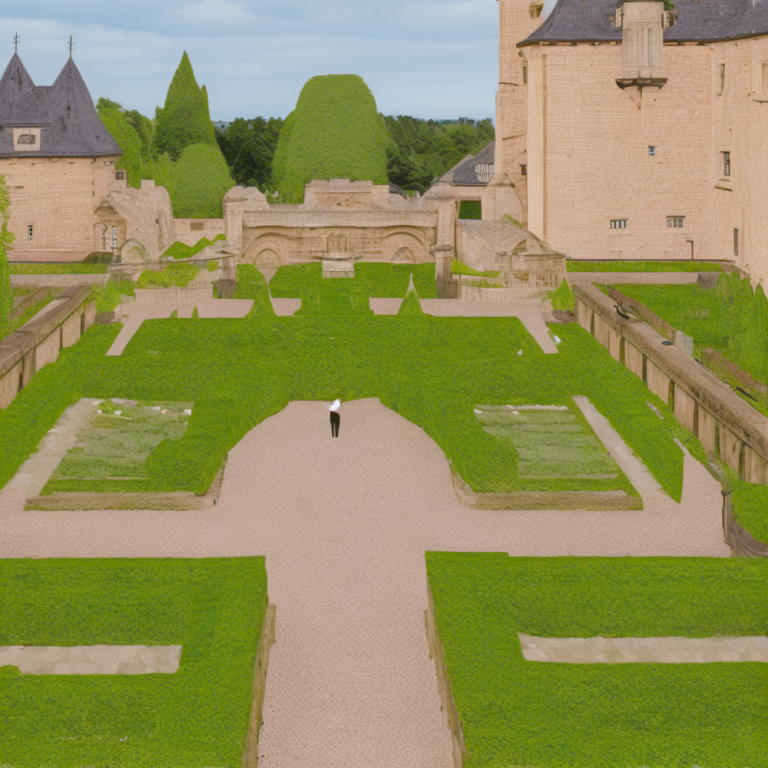}}
        \end{overpic} &
        
        \begin{overpic}[width=0.249\textwidth]{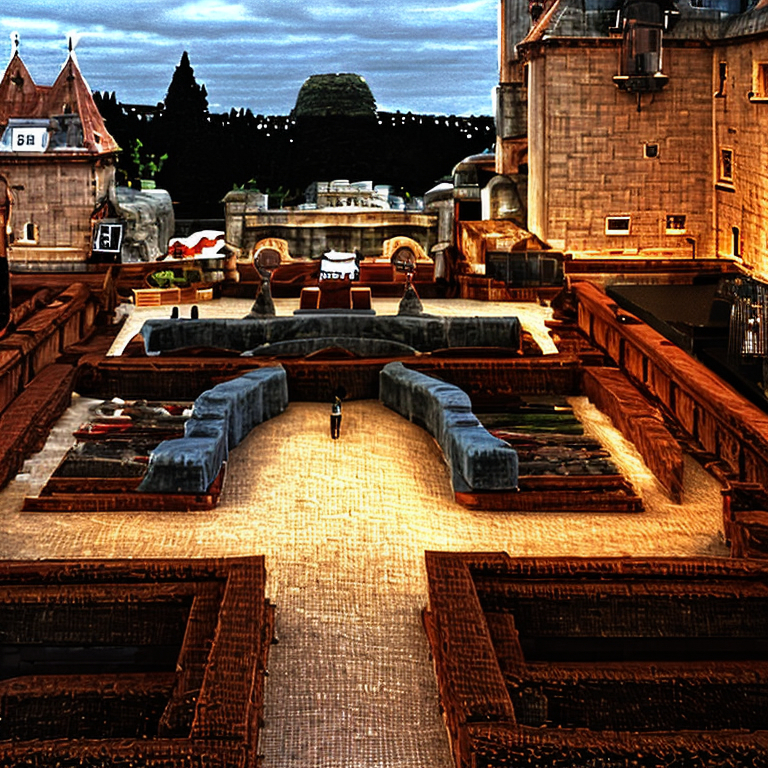}
            \put(0,67){\includegraphics[width=0.08\textwidth]{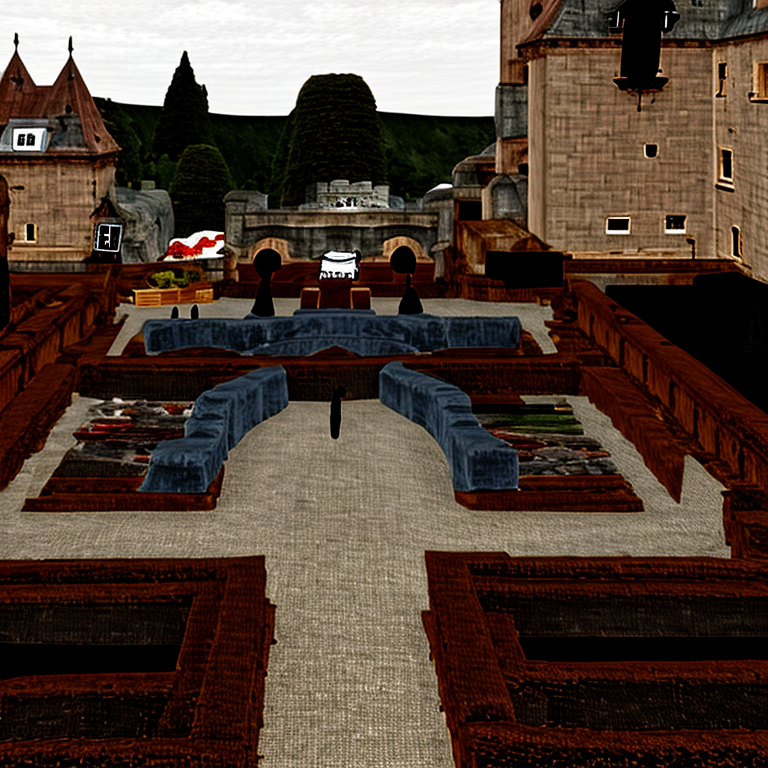}}
        \end{overpic} &
        
        \begin{overpic}[width=0.249\textwidth]{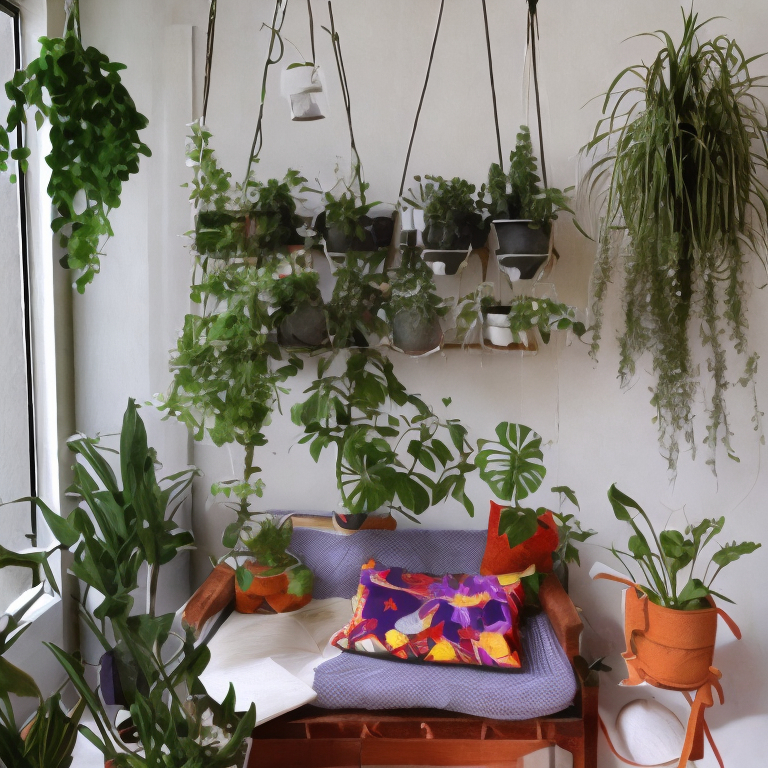}
            \put(0,67){\includegraphics[width=0.08\textwidth]{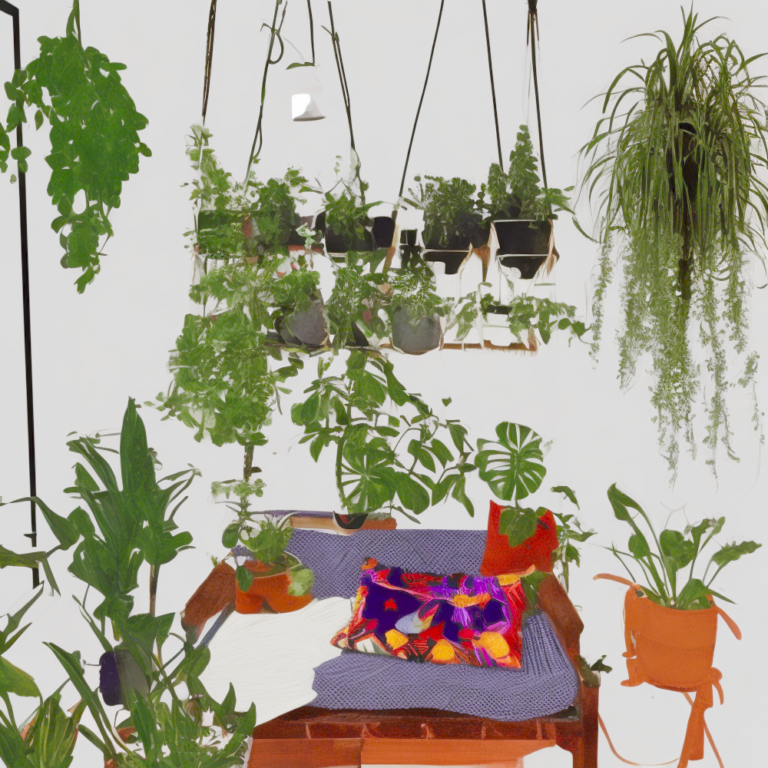}}
        \end{overpic} &
        
        \begin{overpic}[width=0.249\textwidth]{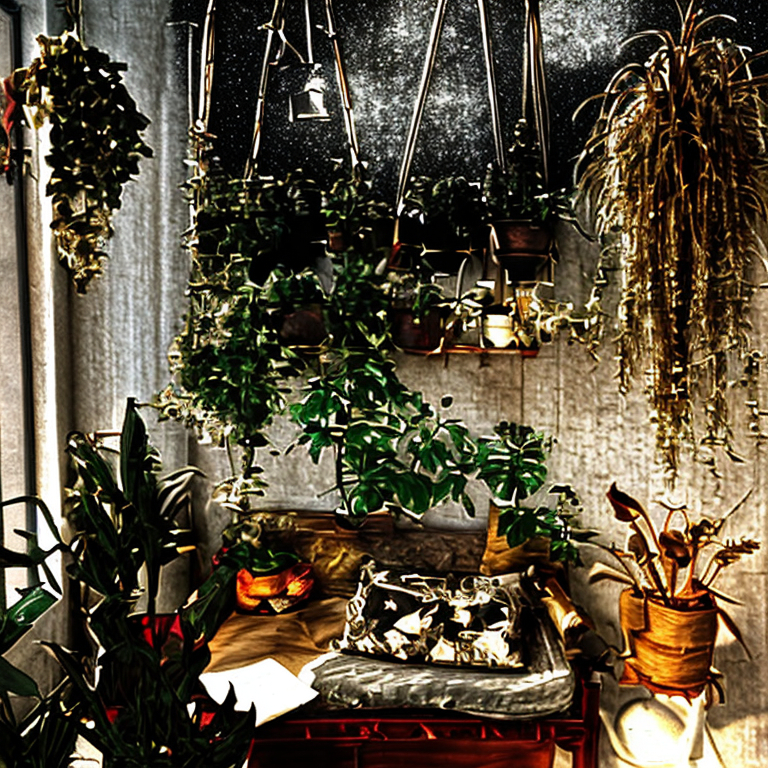}
            \put(0,67){\includegraphics[width=0.08\textwidth]{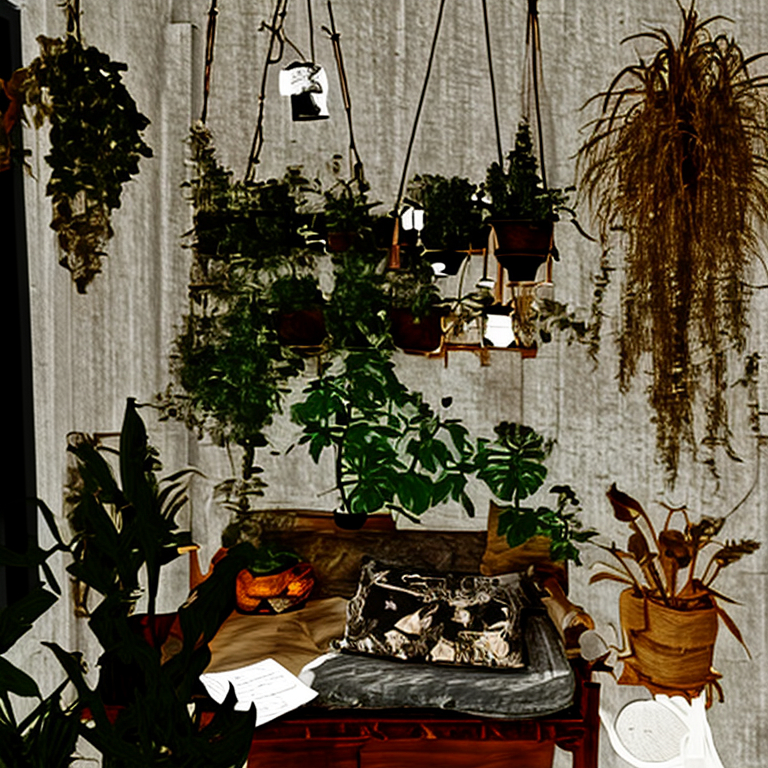}}
        \end{overpic} \\
        
        {\footnotesize Latent-space ControlNet} & {\footnotesize Standard Controlnet}  & 
        {\footnotesize Latent-space ControlNet} & {\footnotesize Standard Controlnet}  \\
    \end{tabular}
    
    \caption{Comparison of G-buffer generation methods across four scenes. Each rendered image includes its corresponding base color representation inset in the corner, providing insight into the material properties generated by each method. Our Latent-space ControlNet approach (left in each pair) consistently produces superior results with more accurate geometric consistency and finer texture details compared to standard approaches (right in each pair). The comparison spans a variety of interior styles from elegant bedrooms to minimalist Japanese-inspired spaces, demonstrating our method's versatility.}
    \label{fig:comparisonlatentcontrolnet}
\end{figure*}

\begin{figure*}[htbp]
    \centering
    \setlength{\tabcolsep}{1pt}
    \begin{tabular}{c c c c c c}

        \textbf{w/o Branch} & \textbf{w/ Branch} & \textbf{Ground truth} & 
        \textbf{w/o Branch} & \textbf{w/ Branch} & \textbf{Ground truth} \\
        
        \includegraphics[width=0.16\textwidth]{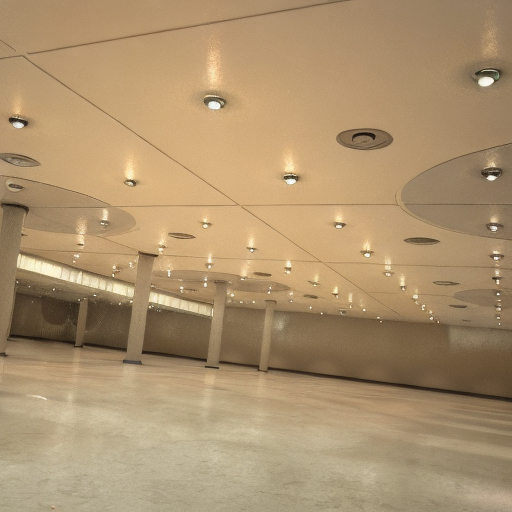} &
        \includegraphics[width=0.16\textwidth]{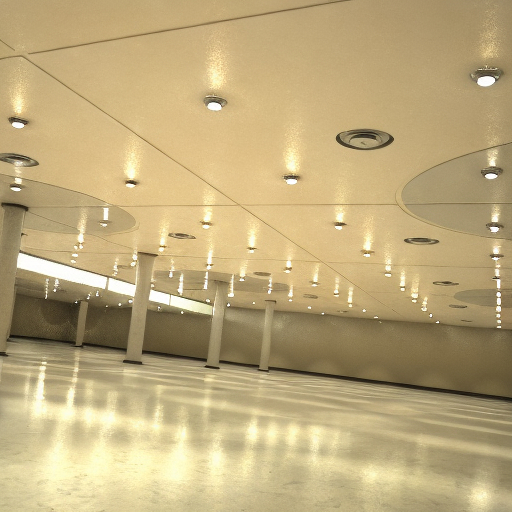} &
        \includegraphics[width=0.16\textwidth]{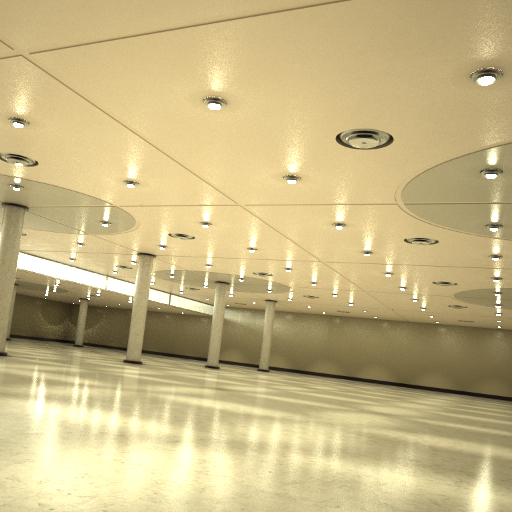} &
        \includegraphics[width=0.16\textwidth]{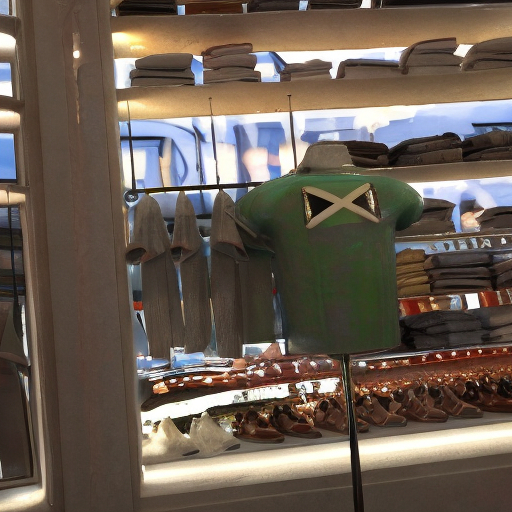} &
        \includegraphics[width=0.16\textwidth]{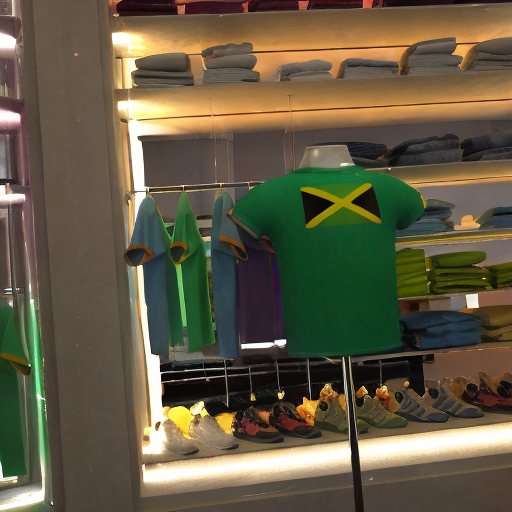} &
        \includegraphics[width=0.16\textwidth]{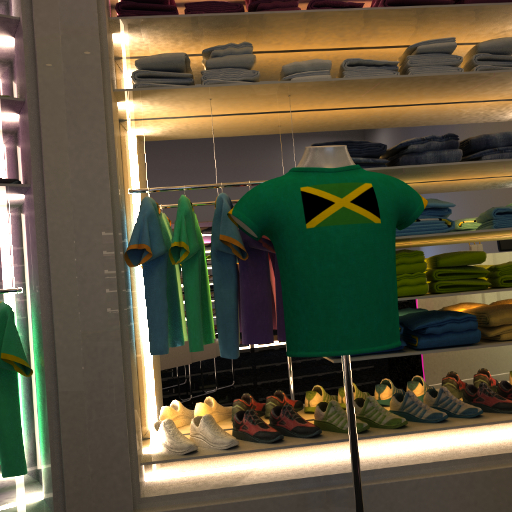} \\
        
        \includegraphics[width=0.16\textwidth]{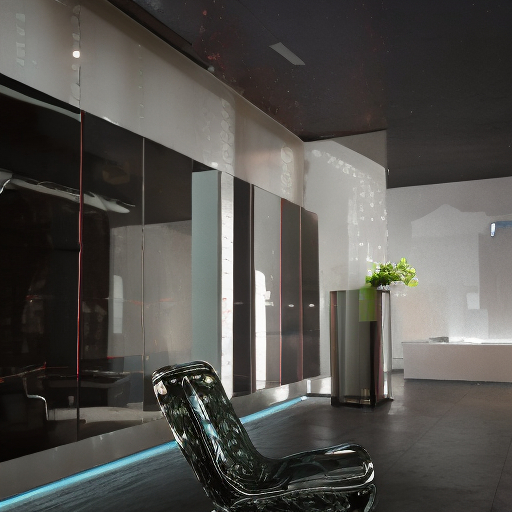} &
        \includegraphics[width=0.16\textwidth]{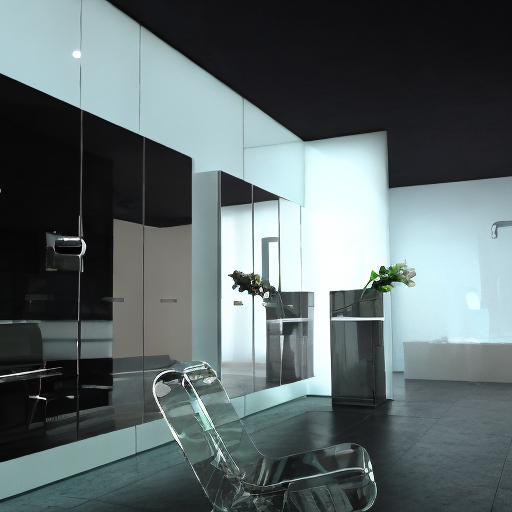} &
        \includegraphics[width=0.16\textwidth]{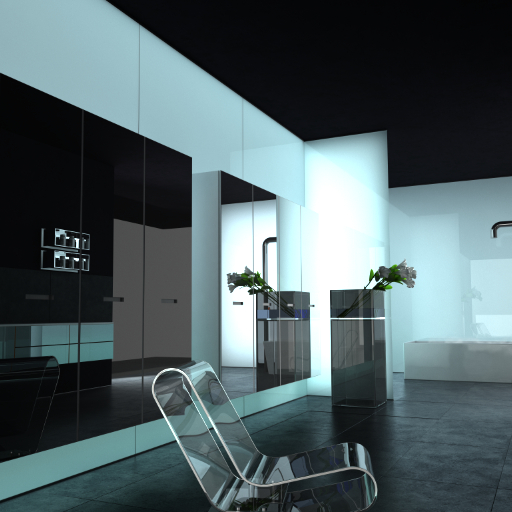} &
        \includegraphics[width=0.16\textwidth]{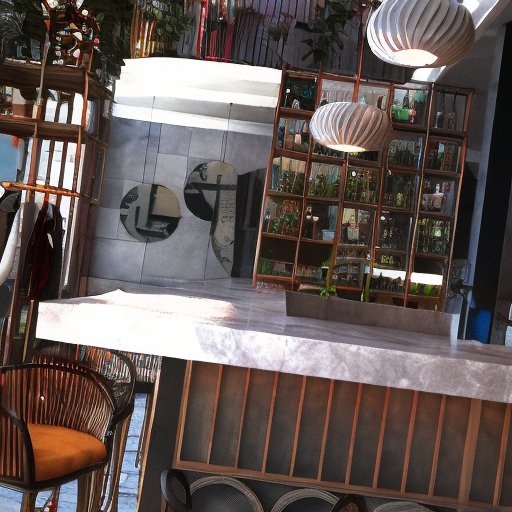} &
        \includegraphics[width=0.16\textwidth]{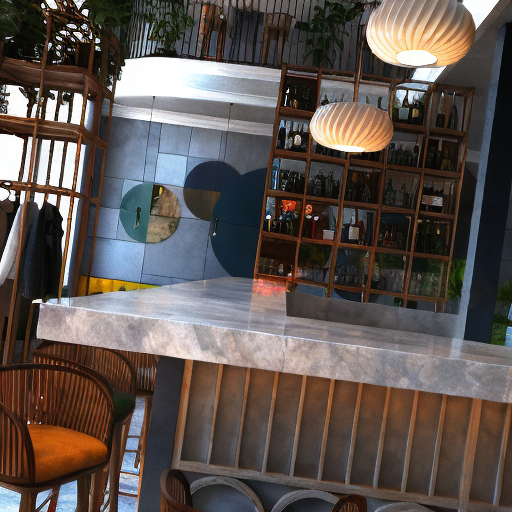} &
        \includegraphics[width=0.16\textwidth]{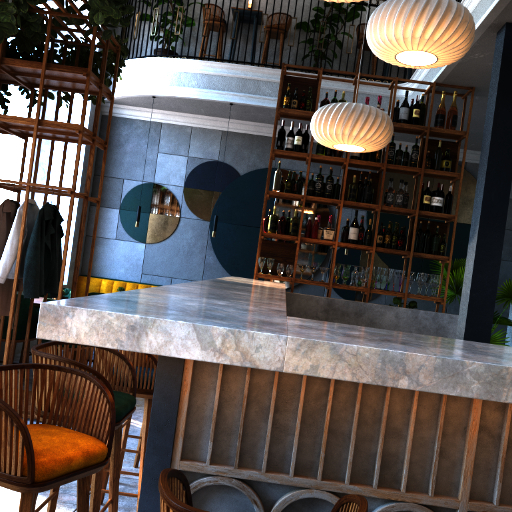} \\
    \end{tabular}
    \caption{Ablation of G-buffer to Final Image with or without Branch Networks. This figure illustrates the impact of Branch Networks on G-buffer rendering. Results show that including Branch Networks produces outputs more closely aligned with the ground truth. All G-buffers and ground-truth images are from the Hypersim dataset.}
    \label{fig:comparisonsubnetwork}
\end{figure*}

\begin{figure*}[htbp]
    \centering
    \setlength{\tabcolsep}{1pt}
    \begin{tabular}{ccccc}
    \includegraphics[width=0.19\textwidth]{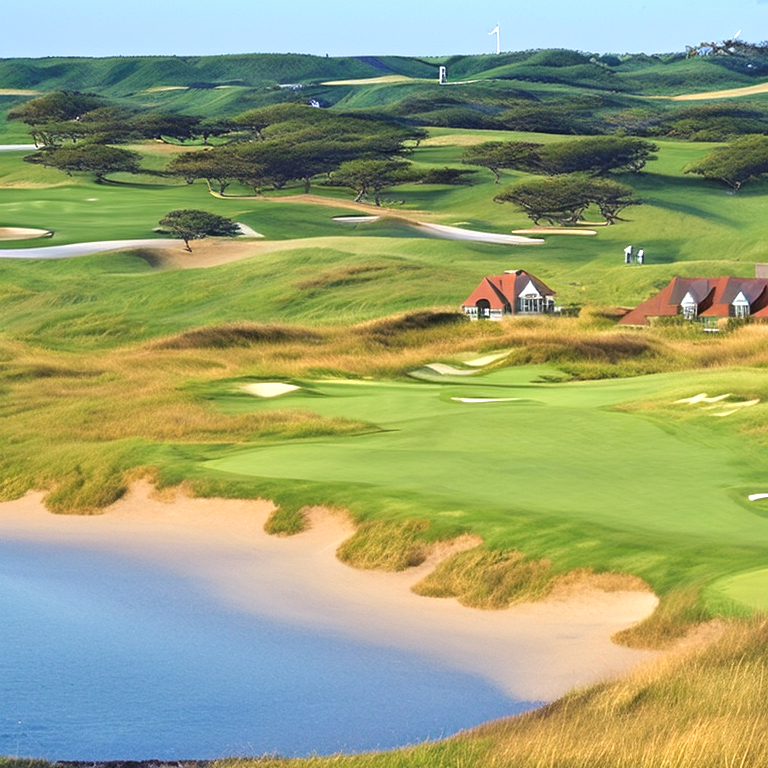} &
    \includegraphics[width=0.19\textwidth]{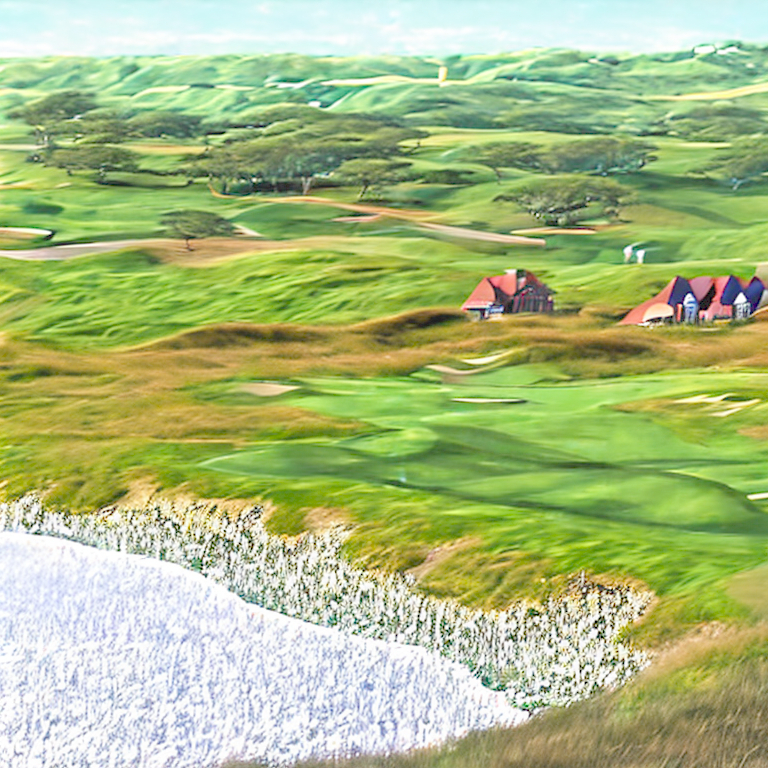} &
    \includegraphics[width=0.19\textwidth]{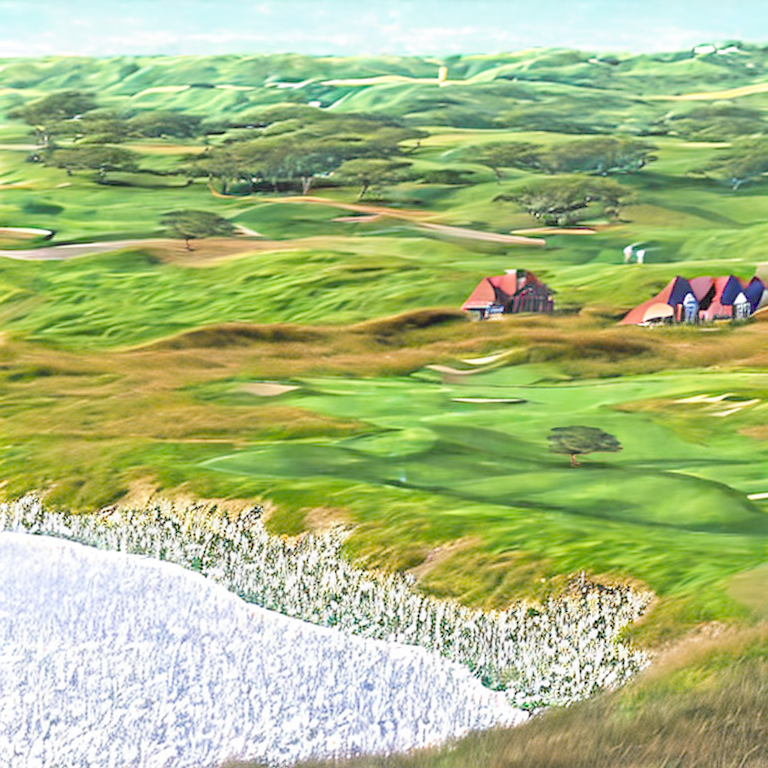}
    & \includegraphics[width=0.19\textwidth]{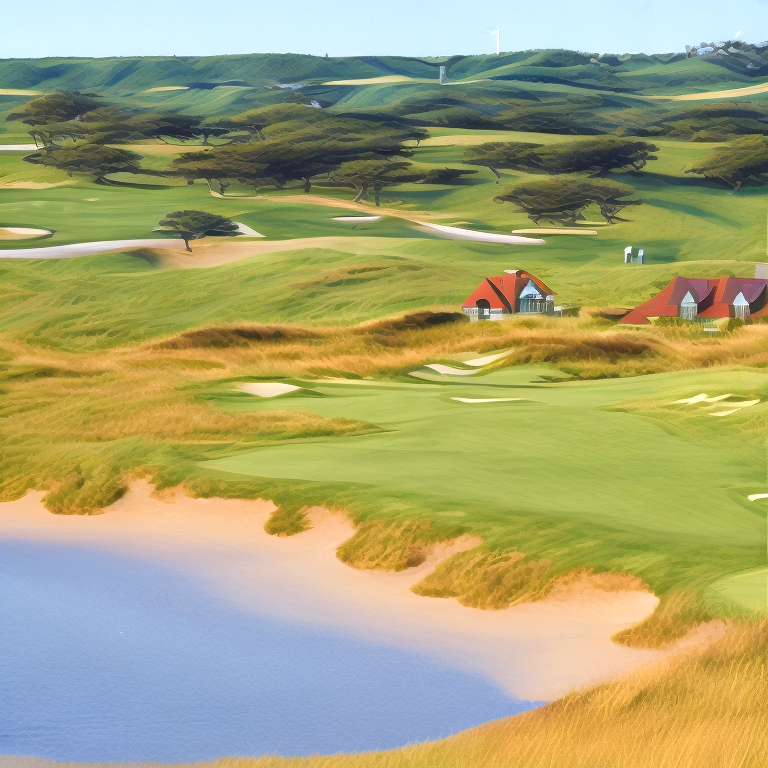} &
    \includegraphics[width=0.19\textwidth]{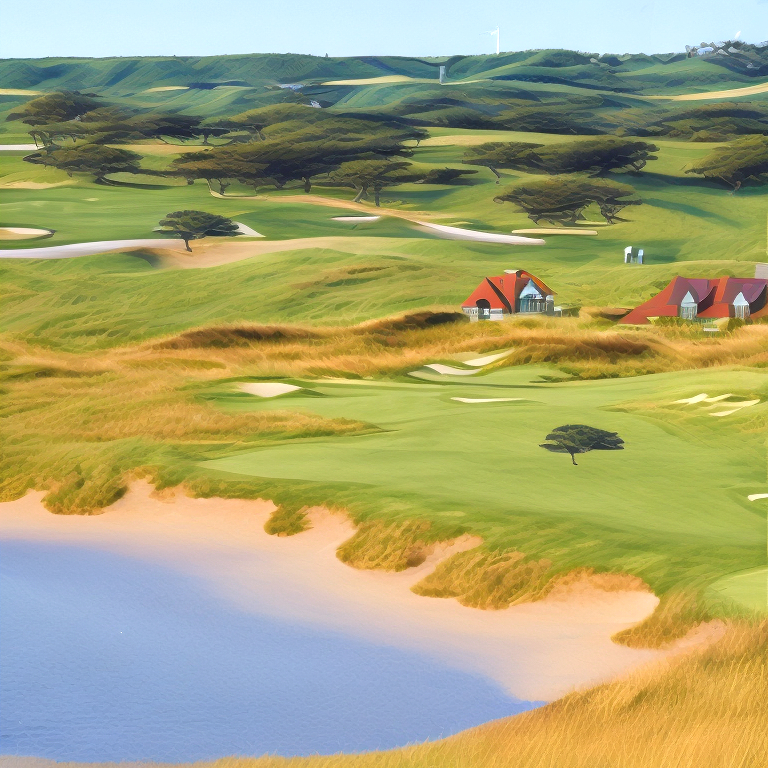} \\
    Stable diffusion & \revtwo{RGB$\leftrightarrow$X} & \revtwo{RGB$\leftrightarrow$X} edited& Ours & Ours edited \\
    \end{tabular}
    \caption{Outdoor scene with lake and houses. Our method  maintains accurate water surface representation and natural integration with the landscape, while \revtwo{RGB$\leftrightarrow$X} produces significant artifacts at the water boundary and fails to correctly render the lake's surface properties compared to the original image.}
    \label{fig:comparisonfullpipe}
\end{figure*}

        


\begin{figure*}[t]
    \centering
    \includegraphics[width=\linewidth]{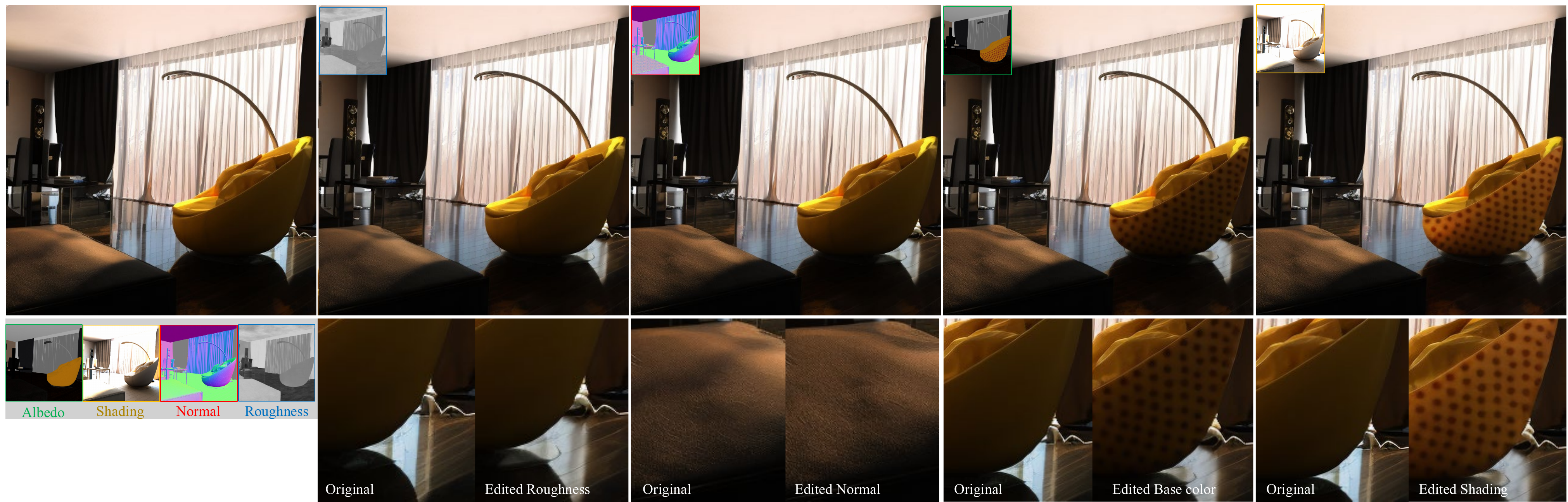}
    \caption{Starting from a base indoor scene and original G-buffer (left), we demonstrate the effects of modifying different G-buffer channels: increasing the floor's roughness to make it more rough, modifying the normal map of the stool in the lower left corner to add multiple small bumps/protrusions, altering the base color/albedo by adding a special pattern to the sofa in the lower right corner, and adjusting shading values to change the lighting effects. This editability is enabled by our method's ability to generate consistent and \rev{PBR-inspired} G-buffers with clean channel separation.}
    \label{fig:gbuffer_editability}
\end{figure*}

\section{Results and Comparisons}
\label{sec:results}

In this section, we present multiple comparison results and ablation studies. Our evaluation includes both objective metrics and user studies, followed by detailed ablations of our architectural choices and comparisons with existing methods on generation and editing tasks.

\subsection{Quantitative Evaluation}
We evaluated our approach on 5,000 samples using multiple metrics: FID~\cite{10.5555/3295222.3295408} for image quality, CLIP Text-Image Score for semantic alignment between images and prompts, and CLIP Aesthetic Score for visual quality, which quantifies visual quality by comparing an image's similarity to positive (e.g., "professional photograph") versus negative descriptors (e.g., "poor composition"). All evaluations use the CLIP ViT-B/32 model.

\begin{table}[htbp]
    \centering
    \caption{Comparison of different methods across objective quality metrics. $\downarrow$ indicates that lower values are better, while $\uparrow$ indicates that higher values are better.}
    \label{tab:quantitative_evaluation}
    \begin{tabular}{lccc}
        \toprule
        \textbf{Method} & \textbf{FID}$\downarrow$ & \textbf{CLIP-T}$\uparrow$ & \textbf{CLIP-A}$\uparrow$ \\
        \midrule
        Stable diffusion & 22.96 & 0.29 & 0.0042 \\
        Ours & 22.97 & 0.29 & 0.0054 \\
        \revtwo{RGB$\leftrightarrow$X} w/ our buffer & 25.86 & 0.28 & -0.0001 \\
        \revtwo{RGB$\leftrightarrow$X} & 49.94 & 0.18 & -0.0021 \\
        \bottomrule
    \end{tabular}
\end{table}

As shown in Table~\ref{tab:quantitative_evaluation}, our method achieves comparable FID scores to the base model, indicating that our additional G-buffer processing and rendering steps maintain image quality. \rev{Note that ``\revtwo{RGB$\leftrightarrow$X} w/ our buffer'' refers to using our Stage-1 network to generate G-buffers, which are then fed into the official \revtwo{RGB$\leftrightarrow$X} renderer for final image synthesis.} For semantic alignment, we maintain the same high CLIP Text-Image Score as the base model (0.29), significantly outperforming \revtwo{RGB$\leftrightarrow$X}. Notably, our approach achieves the highest CLIP Aesthetic Score (0.0054), demonstrating superior visual quality compared to all baselines. 

\textbf{User Study and Analysis.} We conducted a user study to evaluate the perceived quality and realism of our method compared to competing approaches. A total of 156 participants were recruited. Each participant completed 30 forced-choice questions, with two images shown side-by-side in each question (our result vs.\ a baseline result). The participants were instructed to choose which image they felt looked better or more visually plausible. Please check the supplementary material for more details. 
Each participant received the 30 questions in a randomized order to mitigate potential ordering effects. Within each question, the left-right placement of our output vs.\ the baseline output was also randomized to avoid positional bias. We emphasized that participants should pay attention to both local details and global consistency. Participants spent approximately 5-10 minutes finishing all 30 questions.

As shown in Figure~\ref{fig:user_study_bar}, our method consistently outperformed baselines across all evaluation criteria, with preference rates ranging from 64.48\% to 72.67\%. These findings demonstrate the superiority of our approach in producing realistic and visually appealing results for both generation and editing tasks.

\rev{\textbf{G-buffer Quality Validation}
To validate the quality of our generated G-buffers, we evaluate against ground-truth G-buffers from the InteriorVerse test set. Table~\ref{tab:gbuffer_validation} reports per-channel metrics. Note that for text-to-image generation, there is no unique ground-truth G-buffer for a given prompt; these metrics serve as a reference, while our primary validation comes from downstream task performance (Stage-2 reconstruction, user studies, and editing experiments).
}
\begin{table}[htbp]
    \centering
    \caption{\rev{Quantitative evaluation of generated G-buffer quality against ground truth on InteriorVerse.}}
    \label{tab:gbuffer_validation}
    \begin{tabular}{lccccc}
        \toprule
        & \textbf{Albedo} & \textbf{Normal} & \textbf{Depth} & \textbf{Roughness} & \textbf{Metallic} \\
        \midrule
        PSNR$\uparrow$ & 17.6 & 20.3 & 16.3 & 14.3 & 14.7 \\
        LPIPS$\downarrow$ & 0.17 & 0.16 & 0.23 & 0.32 & 0.31 \\
        \bottomrule
    \end{tabular}
\end{table}

\subsection{Ablation Study}

\textbf{Text-to-G-buffer Ablation.}
We conduct an ablation study on different strategies for text-to-G-buffer generation, illustrated in Figure~\ref{fig:comparisoncontrolnet}. To ensure fair comparisons and maintain consistent outputs, all methods use the same random noise, prompt, global seed, and generator. The first row shows an \revtwo{RGB$\leftrightarrow$X}~\cite{zeng2024rgb} network connected to a complete Stable Diffusion~2 pipeline. The second row removes ControlNet and fine-tunes only the Stable Diffusion~2 UNet on our dataset, using the same training duration. The third row employs our proposed Latent ControlNet.

Without shading map, architecture like \revtwo{RGB$\leftrightarrow$X} inpainting network, requiring first rendering result, using a mask to select insertion regions, then blending original and inpainted images.  Otherwise, unintended changes occur. In Figure~\ref{fig:comparisonwithrgbx}, the \revtwo{RGB$\leftrightarrow$X} baseline is blended with a mask, whereas our results are produced without masking. Testing showed without shading map, lighting in most regions differs from original. Shading map serves dual purposes: ensuring stability and enabling lighting editing.

\textbf{Latent ControlNet Ablation.}
To evaluate the efficacy of our proposed latent-space control mechanism, we conducted an ablation study comparing our method with the original ControlNet approach across diverse scenes. As illustrated in Figure \ref{fig:comparisonlatentcontrolnet}, our Latent-space ControlNet implementation (left in each pair) consistently produces higher quality images that more accurately match the input prompts compared to standard approaches (right in each pair). This poor performance of standard ControlNet has also been documented in other related papers \cite{IntrinsicDiffusionluo}. Figure 6 of that paper demonstrates that standard ControlNet leads to albedo color mismatches. The comparative analysis reveals a significant limitation in the original ControlNet method, where mismatches between base color representations and the intended prompt specifications frequently occur, as highlighted in the bolded sections of our caption. These discrepancies are clearly visible in the base color representations inset in each rendered image, demonstrating how standard approaches struggle with prompt-material coherence. Our Latent-space ControlNet successfully addresses this limitation, producing results that faithfully adhere to prompt specifications across a variety of environments, including both interior and exterior scenes. 

\textbf{G-buffer to Final Image With/Without Branch Networks.}
Figure~\ref{fig:comparisonsubnetwork} compares our branch-based rendering (employing sub-networks $\mathcal{G}$, $\mathcal{M}$, $\mathcal{L}$, and a merge module $\mathcal{H}$) against a single ControlNet trained end-to-end without such factorization. For clarity, we do not use the entire pipeline; instead, we take G-buffers directly from the dataset and feed them into the second-stage network to render images, allowing a direct comparison with ground truth.

Our \rev{PBR-inspired} branch approach captures lighting, geometry, and material properties more consistently. In contrast, direct ControlNet training exhibits mild color mismatches (e.g., background color artifacts) and struggles with complex lighting effects (e.g., ground reflections). Transparent or highly reflective objects (glass seats, mirrors, metallic surfaces) are also rendered more accurately by our branched model. The single ControlNet baseline frequently produces metallic reflections in glass objects or overly diffuse reflections on metallic surfaces, undermining realism.
Table \ref{tab:ablation_study} illustrates the performance improvements achieved by incorporating branch networks into our model. Specifically, the model with branch networks exhibits a significant reduction in Mean Squared Error (MSE) and Learned Perceptual Image Patch Similarity (LPIPS) scores, alongside enhancements in Structural Similarity Index Measure (SSIM) and Peak Signal-to-Noise Ratio (PSNR) metrics, compared to the model without branch networks.
\begin{table}[htbp]
    \centering
    \caption{Performance Comparing Models With and Without Branch Networks. $\downarrow$ indicates that lower values are better, while $\uparrow$ indicates that higher values are better.}
    \label{tab:ablation_study}
    \begin{tabular}{lcc}
        \toprule
        \textbf{Metric} & \textbf{w/o Branch Networks} & \textbf{w/ Branch Networks} \\
        \midrule
        MSE $\downarrow$  & 0.0288  & 0.0068 \\
        LPIPS $\downarrow$ & 0.2686  & 0.0973 \\
        SSIM $\uparrow$ & 0.6350  & 0.8072 \\
        PSNR $\uparrow$ & 15.9983 & 21.9883 \\
        \bottomrule
    \end{tabular}
\end{table}
\subsection{Comparison with Related Work}

\textbf{Insertion comparison with \revtwo{RGB$\leftrightarrow$X}}
Figure~\ref{fig:comparisonwithrgbx} compares our method and \revtwo{RGB$\leftrightarrow$X} \cite{zeng2024rgb} on insertion tasks. To ensure fairness, we use \emph{existing} G-buffers from the Hypersim dataset. The left column shows the original image, the middle column shows our result, and the right column is \revtwo{RGB$\leftrightarrow$X}’s output.

Our method produces more natural shadows and lighting for inserted objects. By contrast, \revtwo{RGB$\leftrightarrow$X} often exhibits unnatural artifacts or inaccurate color for inserted objects, especially for reflective or refractive surfaces like glass (e.g., the wine bottle in the final scene). In the second example, where a wooden stump is inserted, our method preserves refraction effects, yielding a coherent scene. \revtwo{RGB$\leftrightarrow$X} fails to capture these details, leading to visually inconsistent results. In the third example, \revtwo{RGB$\leftrightarrow$X} generates odd shadows, whereas ours maintains shape details and realistic shadow casting.

\textbf{Comparison whole Pipeline with \revtwo{RGB$\leftrightarrow$X}}
Figure~\ref{fig:comparisonfullpipe} presents a further comparison between our approach and \revtwo{RGB$\leftrightarrow$X}, highlighting fundamental differences in processing methodology. In this experiment, we use Stable Diffusion outputs processed through VAE networks as inputs for \revtwo{RGB$\leftrightarrow$X}, while our method directly operates on latent representations. The results demonstrate that \revtwo{RGB$\leftrightarrow$X} produces outputs with noticeable artifacts, particularly in outdoor scenes, where the generated images deviate significantly from the original inputs. \rev{This may occur because \revtwo{RGB$\leftrightarrow$X} requires an additional decode–encode cycle, whereas our method operates entirely in latent space.} Consequently, edited results from \revtwo{RGB$\leftrightarrow$X} exhibit these artifacts too. In contrast, our latent-based approach produces more coherent and visually consistent results, validating the effectiveness of our direct latent manipulation strategy.

\begin{figure}[htbp]
    \centering
    \setlength{\tabcolsep}{1pt}
    \begin{tabular}{ccc}
        \includegraphics[width=0.32\linewidth]{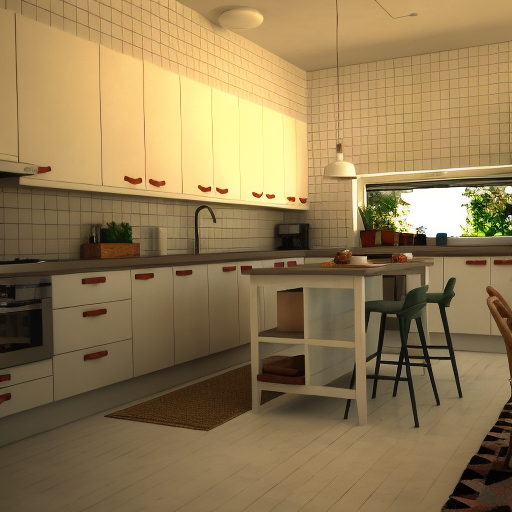} &
        \includegraphics[width=0.32\linewidth]{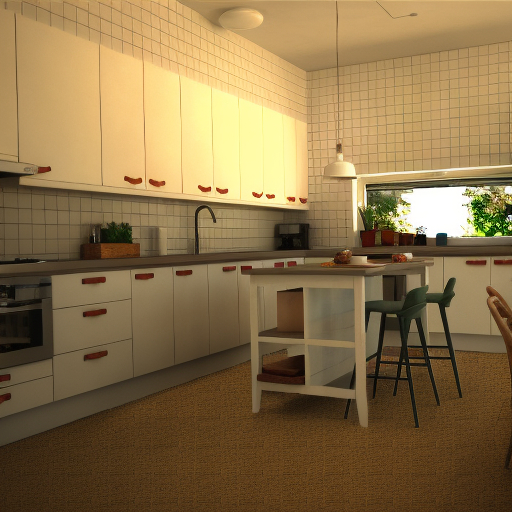} &
        \includegraphics[width=0.32\linewidth]{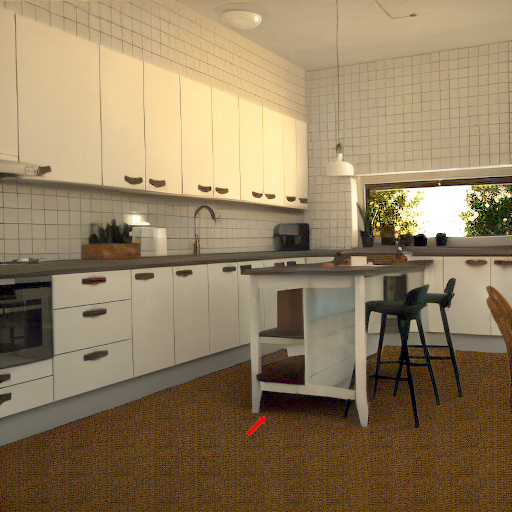} \\

        \includegraphics[width=0.32\linewidth]{image/edit2/imageori.jpg} &
        \includegraphics[width=0.32\linewidth]{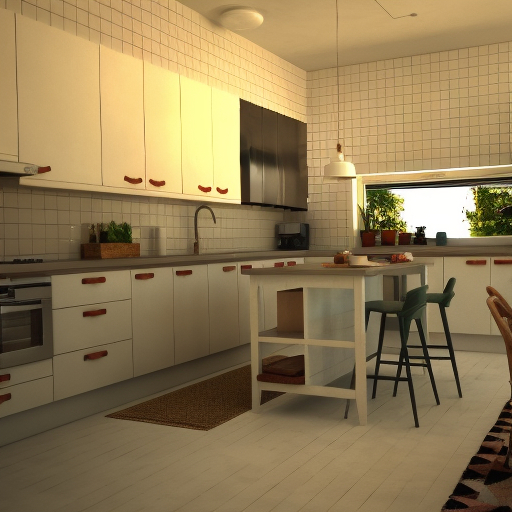} &
        \includegraphics[width=0.32\linewidth]{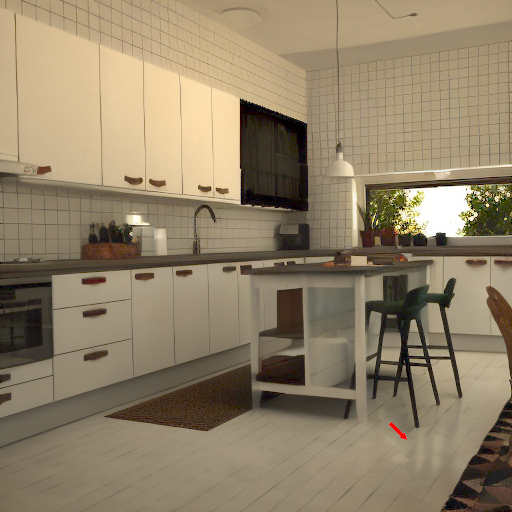} \\
        
        \includegraphics[width=0.32\linewidth]{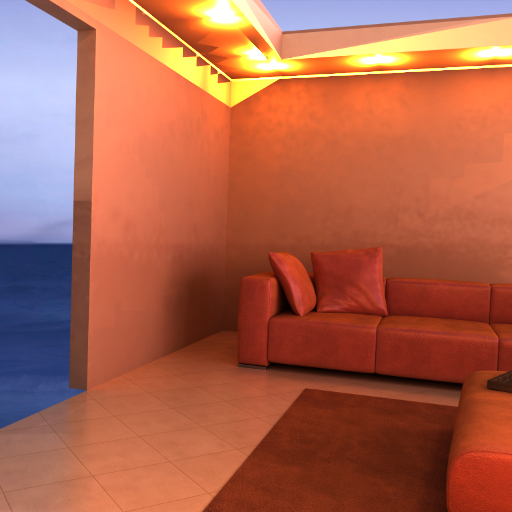} &
        \includegraphics[width=0.32\linewidth]{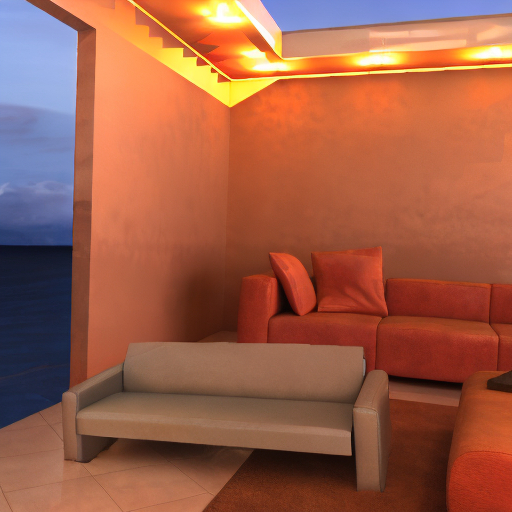} &
        \includegraphics[width=0.32\linewidth]{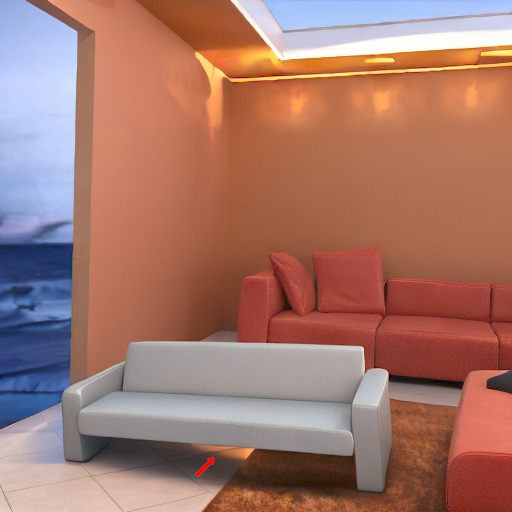} \\

        \includegraphics[width=0.32\linewidth]{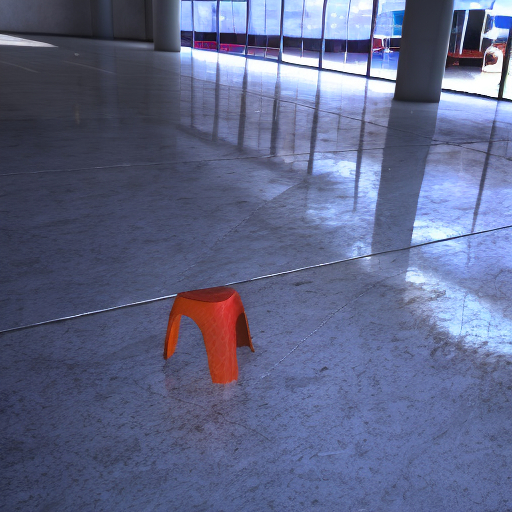} &
        \includegraphics[width=0.32\linewidth]{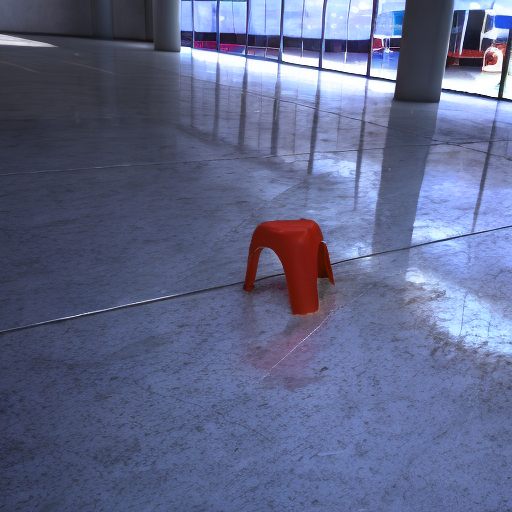} &
        \includegraphics[width=0.32\linewidth]{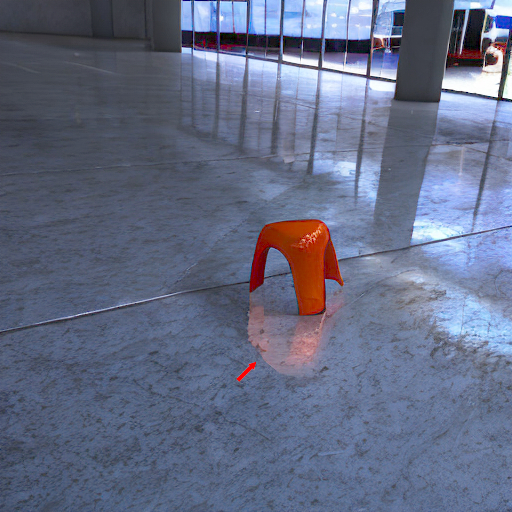} \\        
        
        Original & Ours & IntrinsicEdit \\
    \end{tabular}
    \caption{\rev{Qualitative comparison with IntrinsicEdit~\cite{10.1145/3731173} on editing tasks.} \revtwo{In the first two indoor examples, IntrinsicEdit exhibits noticeable color shifts from the original images. Specifically, in the first example, the shadow position (under the table) changes unexpectedly; and reflections of the stool appear on the floor (not present in the original) in the second. In the third example, bright regions appear beneath the sofa, similar to the effect observed with RGB$\leftrightarrow$X (see Figure~\ref{fig:comparisonwithrgbx}). In the fourth example, the chair's reflection on the marble surface differs from the original.}}
    \label{fig:comparison_intrinsicedit}
\end{figure}

\rev{\textbf{Comparison with IntrinsicEdit.}
Figure~\ref{fig:comparison_intrinsicedit} presents a qualitative comparison with IntrinsicEdit~\cite{10.1145/3731173} on editing tasks.} \revtwo{IntrinsicEdit is a training-free 
optimization framework built on top of RGB$\leftrightarrow$X, which offers 
flexibility but may also inherit limitations from its base model. In the first 
two indoor examples, IntrinsicEdit exhibits color shifts from the 
original images: specifically, the shadow position under the table changes 
unexpectedly in the first example, and reflections of the stool appear 
on the floor (not present in the original) in the second. In the third example 
(sofa insertion), IntrinsicEdit produces bright regions beneath the sofa, 
similar to the effect observed with RGB$\leftrightarrow$X (see 
Figure~\ref{fig:comparisonwithrgbx}), suggesting this issue may stem from the 
underlying base model. In the fourth example, the chair's reflection on the 
marble surface differs from the original.}

\textbf{Comparison Moving Object with Diffusion Handles.}
Figure~\ref{fig:comparisonmove} compares our method to a diffusion-based editing approach \cite{pandey2024diffusion} that specializes in moving objects within an image. We generated multiple outputs (over five) for the competing method and chose its best result for display; even so, it often distorts the background or alters object geometry. For instance, in the first image, the sofa becomes deformed when the chair is moved. In the second image, a background scarf becomes partially transparent, and in the third example, the glass object fails to maintain realistic lighting. The fourth image consistently shows unnatural floor lighting. In contrast, our approach preserves background details, refraction, shadow consistency, and object geometry across all examples, resulting in more reliable and practical edited outputs.

\subsection{Additional Editing Results}

\textbf{Material and Texture Editing.} \rev{As shown in Figure~\ref{fig:comparison_intrinsicedit} (Rows 1-2)}, our method demonstrates multiple material and texture editing capabilities. Starting from an original indoor scene, we expand carpet texture to cover the entire floor surface and transform a wooden cabinet to a metallic finish. These edits are cumulative—each edit includes all previous modifications. These edits highlight our approach's effectiveness in texture synthesis and material property modification while preserving unmodified scene elements.

\textbf{G-buffer Channel Manipulation
} Figure~\ref{fig:gbuffer_editability} illustrates our method's control over individual G-buffer channels. From a base indoor scene, we modify different channels: increasing floor roughness, adding bumps to the stool's normal map, applying a special pattern to the sofa's base color, and adjusting shading values. This editability is enabled by our approach's generation of consistent and \rev{PBR-inspired} G-buffers with clean channel separation. Figure~\ref{fig:teaser} shows object insertion alongside various per-channel modifications and their combined effects.

\section{Discussion and Conclusion}

\textbf{Limitations and Future Work.}
While our method demonstrates strong generalization from indoor training data to outdoor scenes, it may encounter challenges with extremely complex outdoor environments that differ significantly from typical scenes (e.g., underwater environments, aerial landscapes, or abstract artistic compositions). \rev{Our single-bounce approximation and training data bias can result in soft shadows and blurry specular highlights in some cases, as the model does not explicitly handle multi-bounce light transport or high-frequency specular reflections. While outdoor scene quality is improved compared to \revtwo{RGB$\leftrightarrow$X} (see Figure~\ref{fig:comparisonfullpipe} and supplementary materials), results remain weaker than for indoor scenes due to limited outdoor training data.} Additionally, the two-stage pipeline increases inference time compared to direct text-to-image generation. Future work could explore more diverse training datasets and investigate single-stage architectures that maintain our editing capabilities while improving efficiency.

\textbf{Conclusion.}
We presented \rev{a controllable diffusion pipeline for G-buffer generation and rendering}, bridging generative diffusion with fine-grained \rev{scene} control, addressing the fundamental limitation of current text-to-image systems. Our Latent ControlNet operates directly in latent space, avoiding information loss. The \rev{PBR-inspired branch network}, motivated by the rendering equation, achieves 76\% reduction in MSE error while enabling intuitive per-channel editing. Our pipeline design and progressive training strategy successfully learn from limited data while preserving the generative capabilities of models. Extensive evaluations, including user studies with 156 participants, confirm our method's superiority in both generation quality and editing capabilities. This integration enables controllable, \rev{PBR-inspired} image synthesis with channel-wise edits to geometry, materials, and lighting.




\printbibliography

%% file: ablationcontrolnet.tex
\begin{figure}[ht]
    \centering
    \setlength{\tabcolsep}{0.5pt}
    \begin{tabular}{c c c}
        \textbf{RGBX} &
        \textbf{Finetune SD} &
        \textbf{Latent ControlNet} \\
        \begin{overpic}[width=0.16\textwidth]{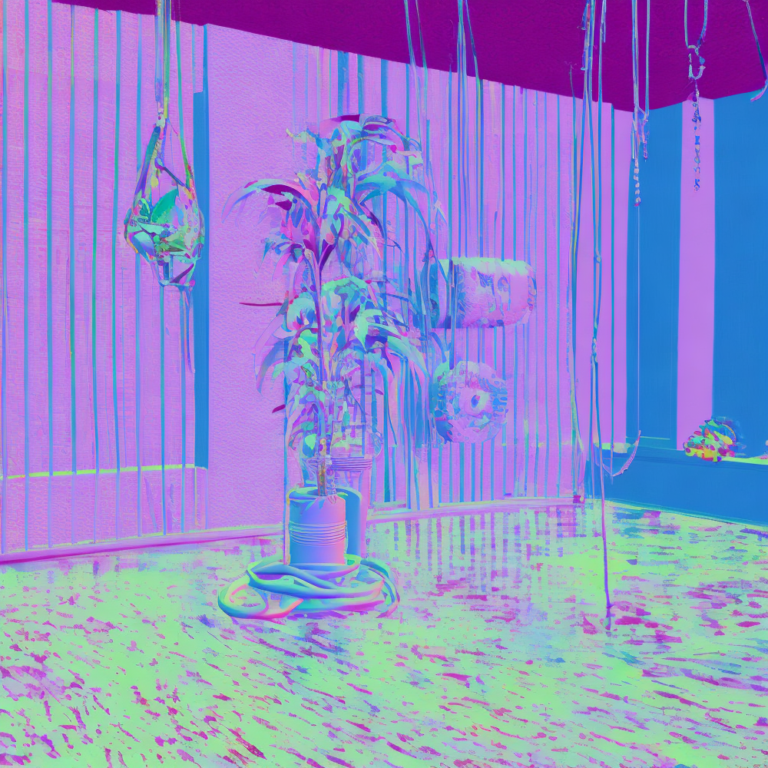}
            \put(0,0){\includegraphics[width=0.04\textwidth]{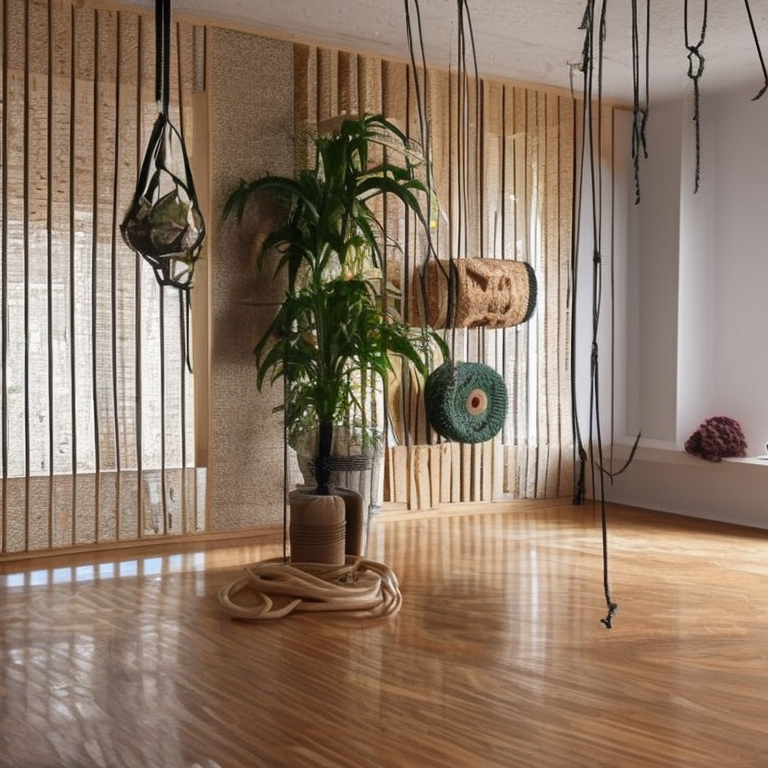}}
        \end{overpic}         &\includegraphics[width=0.16\textwidth]{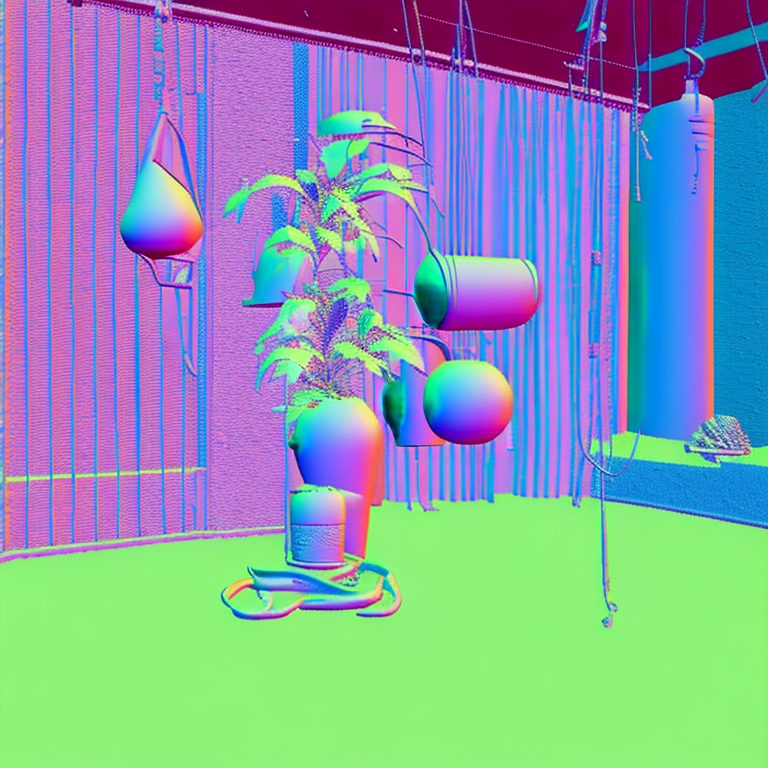}        &\includegraphics[width=0.16\textwidth]{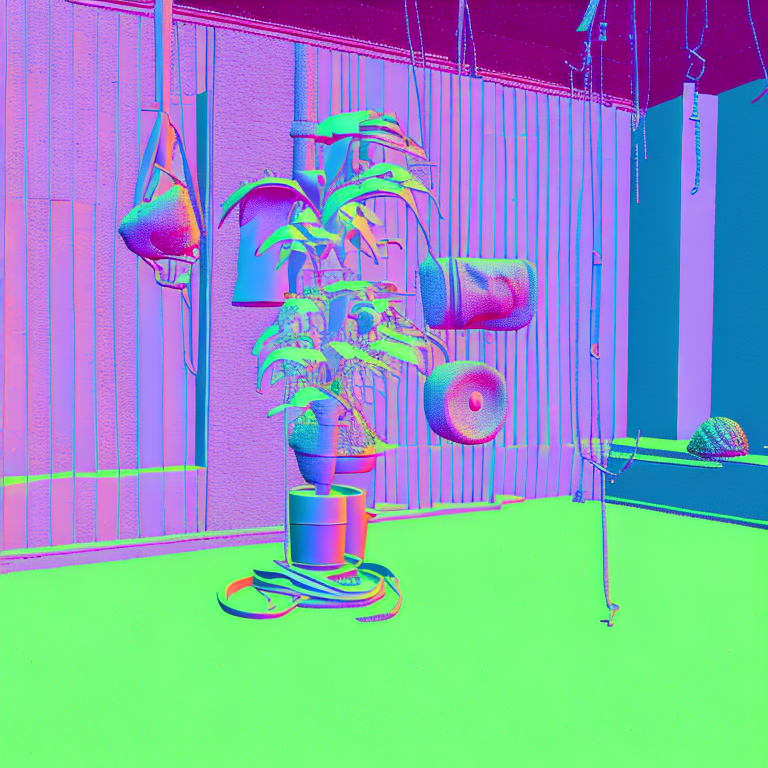}  \\
        \begin{overpic}[width=0.16\textwidth]{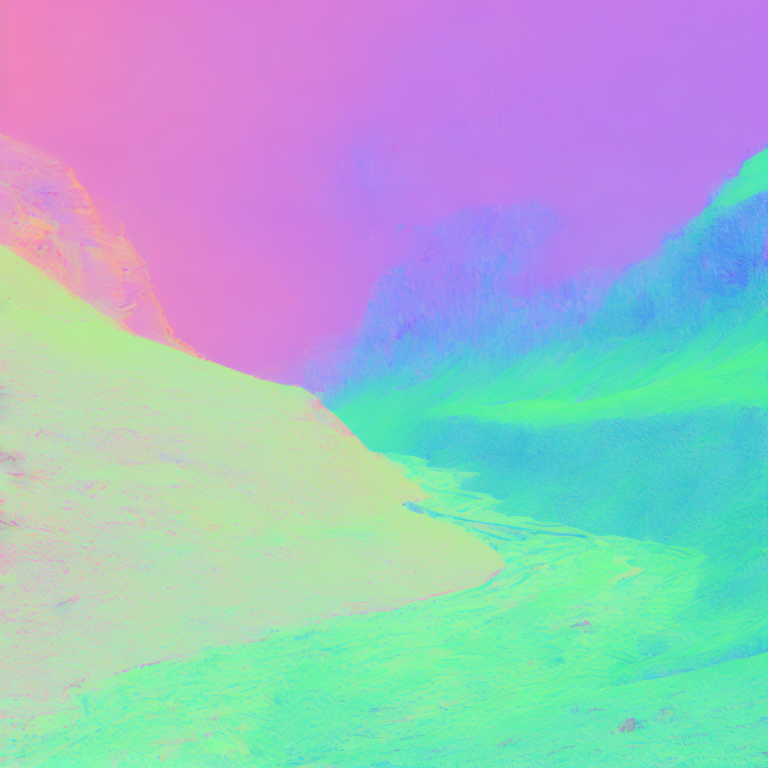}
            \put(0,0){\includegraphics[width=0.04\textwidth]{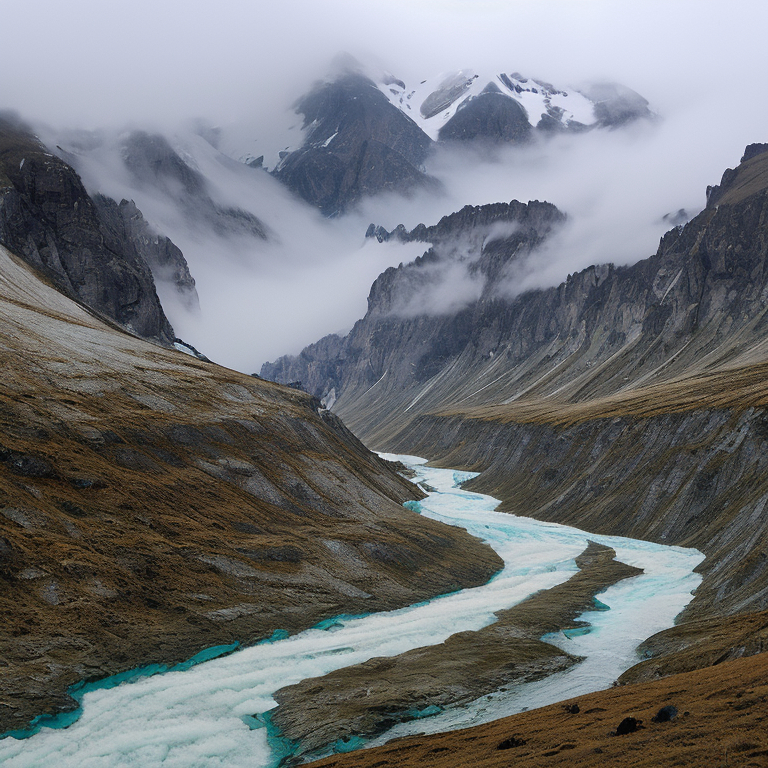}}
        \end{overpic}   &\includegraphics[width=0.16\textwidth]{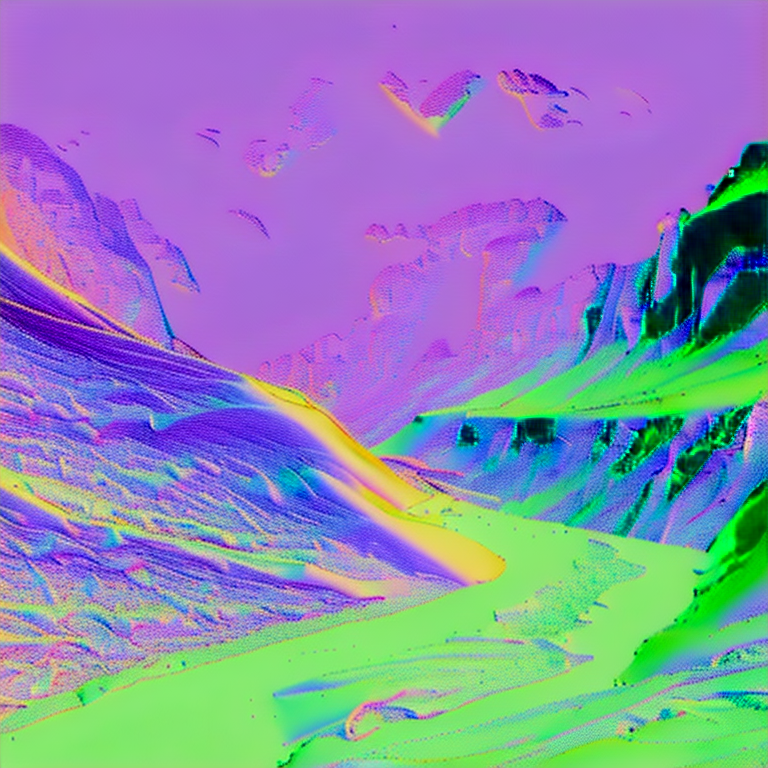}       &\includegraphics[width=0.16\textwidth]{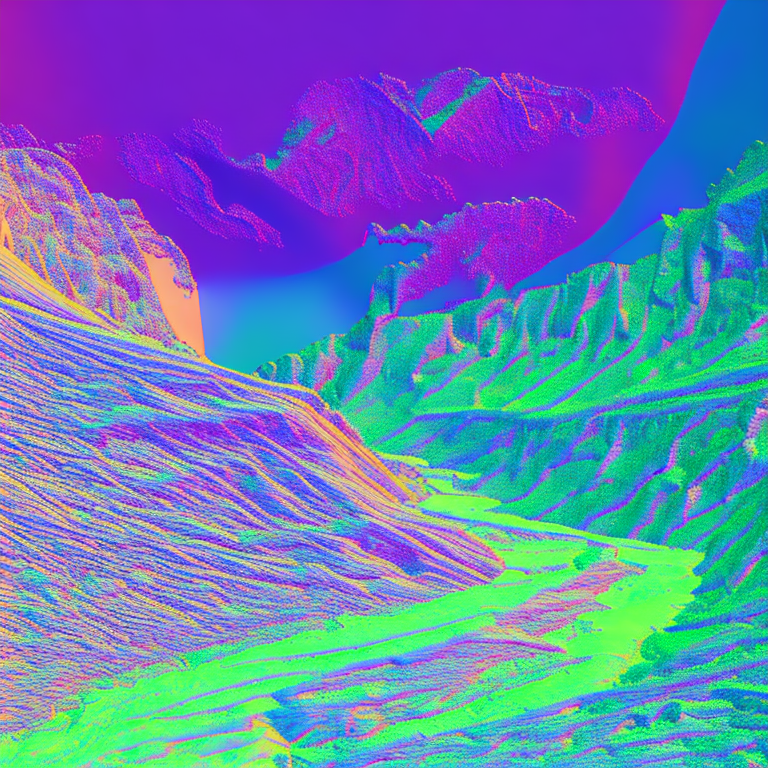}  \\
        \begin{overpic}[width=0.16\textwidth]{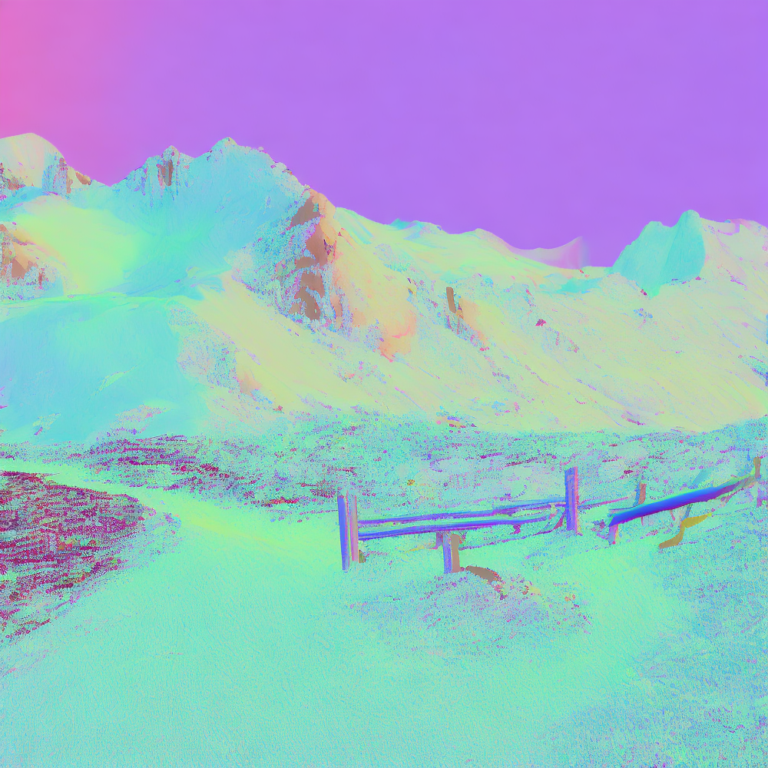}
            \put(0,0){\includegraphics[width=0.04\textwidth]{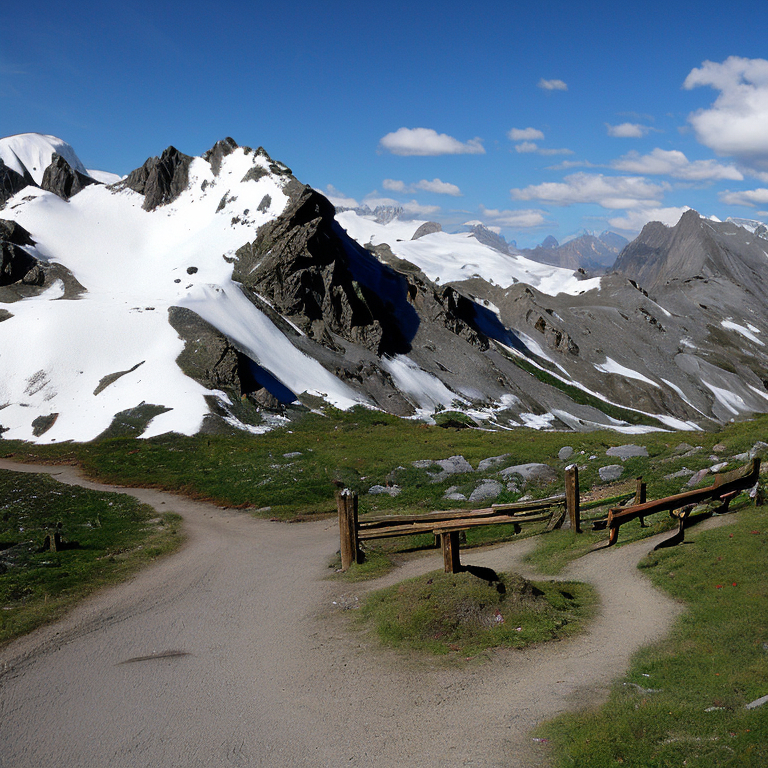}}
        \end{overpic}   &\includegraphics[width=0.16\textwidth]{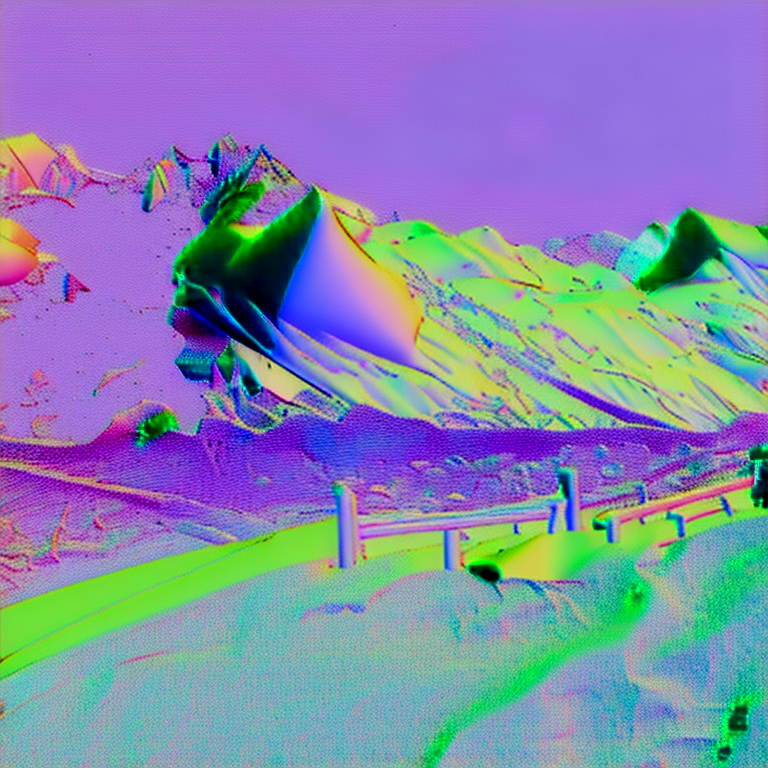}       &\includegraphics[width=0.16\textwidth]{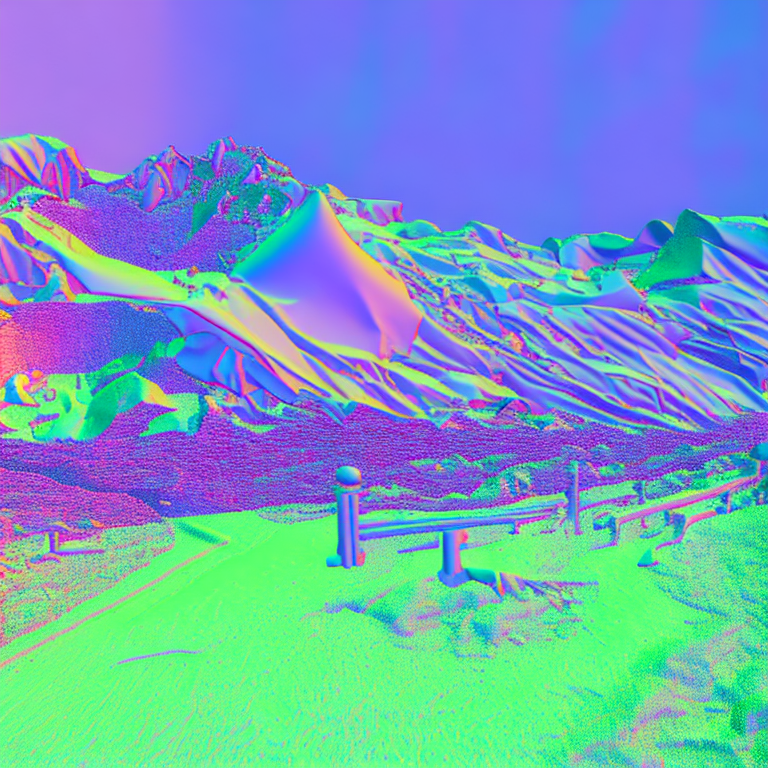} \\
    \end{tabular}
    \caption{Text-to-G-buffer Ablation. This figure compares the performance of three text-to-G-buffer generation approaches across three example scenes (columns), with all images depicting normal maps. The first column shows results from linking the \revtwo{RGB$\leftrightarrow$X} network to the full Stable Diffusion pipeline, using the same noise, seed, and generator as our method. The second column presents outcomes from directly training the Stable Diffusion UNet without ControlNet. The third column showcases results from our full method, demonstrating its superior performance compared to the alternatives.}
    \label{fig:comparisoncontrolnet}
\end{figure}

%% file: comparediffusionhandles.tex
\begin{figure}[htbp]
\centering
\setlength{\tabcolsep}{1pt}
\begin{tabular}{c c c}
\textbf{Original image} & \textbf{Ours} & \textbf{Diffusion Handles} \\
\begin{tikzpicture}
\node[anchor=south west,inner sep=0] (img4) at (0,0)
{\includegraphics[width=0.16\textwidth]{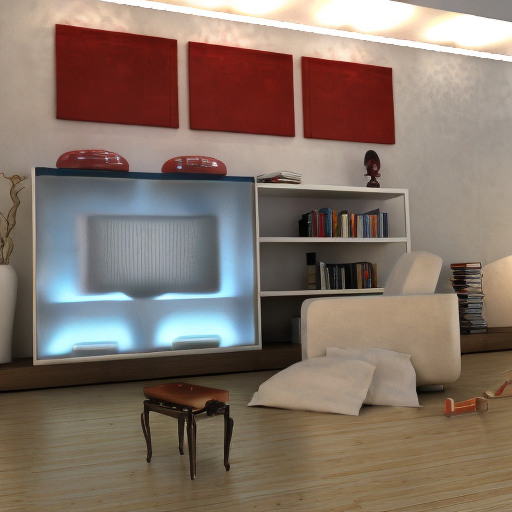}};
\begin{scope}[x={(img4.south east)}, y={(img4.north west)}]
\draw[gray!30,, thick] (0.25,0.05) rectangle (0.5,0.3);
\end{scope}
\end{tikzpicture}
&
\begin{tikzpicture}
\node[anchor=south west,inner sep=0] (img5) at (0,0)
{\includegraphics[width=0.16\textwidth]{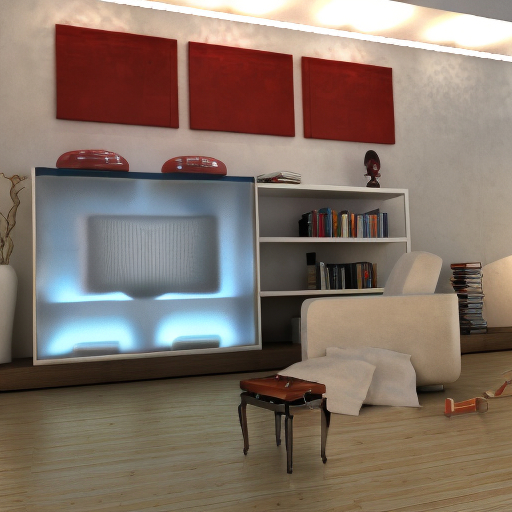}};
\begin{scope}[x={(img5.south east)}, y={(img5.north west)}]
\draw[gray!30,, thick] (0.43,0.05) rectangle (0.68,0.3);
\end{scope}
\end{tikzpicture}
&
\begin{tikzpicture}
\node[anchor=south west,inner sep=0] (img6) at (0,0)
{\includegraphics[width=0.16\textwidth]{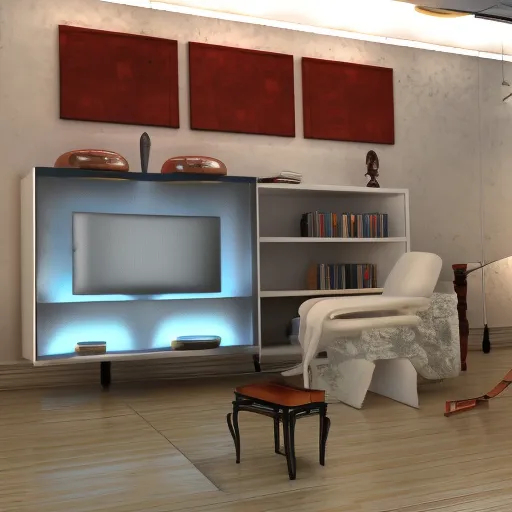}};
\begin{scope}[x={(img6.south east)}, y={(img6.north west)}]
\draw[gray!30,, thick] (0.43,0.05) rectangle (0.68,0.3);
\end{scope}
\end{tikzpicture}
\\
\begin{tikzpicture}
\node[anchor=south west,inner sep=0] (img7) at (0,0)
{\includegraphics[width=0.16\textwidth]{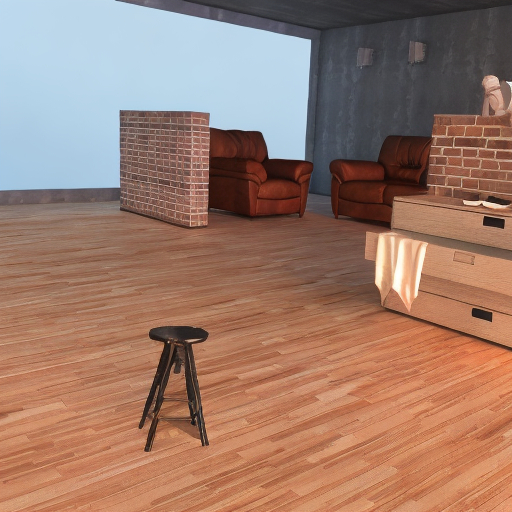}};
\begin{scope}[x={(img7.south east)}, y={(img7.north west)}]
\draw[gray!30,, thick] (0.25,0.1) rectangle (0.45,0.4);
\end{scope}
\end{tikzpicture}
&
\begin{tikzpicture}
\node[anchor=south west,inner sep=0] (img8) at (0,0)
{\includegraphics[width=0.16\textwidth]{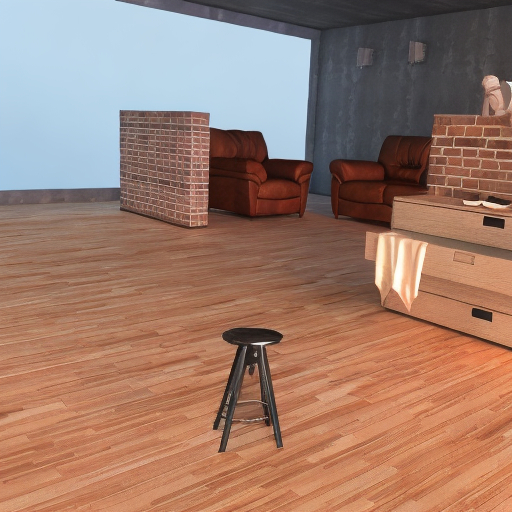}};
\begin{scope}[x={(img8.south east)}, y={(img8.north west)}]
\draw[gray!30,, thick] (0.4,0.1) rectangle (0.6,0.4);
\end{scope}
\end{tikzpicture}
&
\begin{tikzpicture}
\node[anchor=south west,inner sep=0] (img9) at (0,0)
{\includegraphics[width=0.16\textwidth]{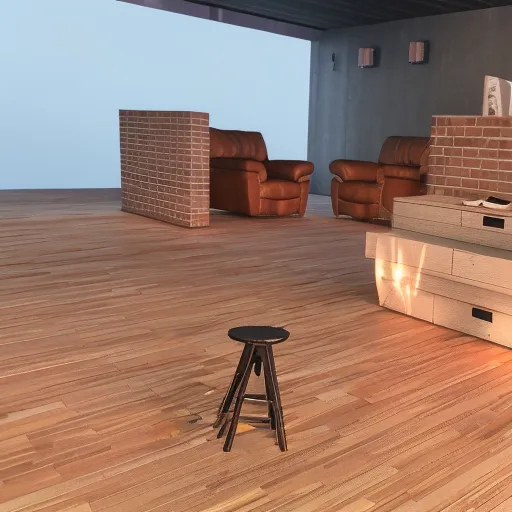}};
\begin{scope}[x={(img9.south east)}, y={(img9.north west)}]
\draw[gray!30,, thick] (0.4,0.1) rectangle (0.6,0.4);
\end{scope}
\end{tikzpicture}
\\
\begin{tikzpicture}
\node[anchor=south west,inner sep=0] (img10) at (0,0)
{\includegraphics[width=0.16\textwidth]{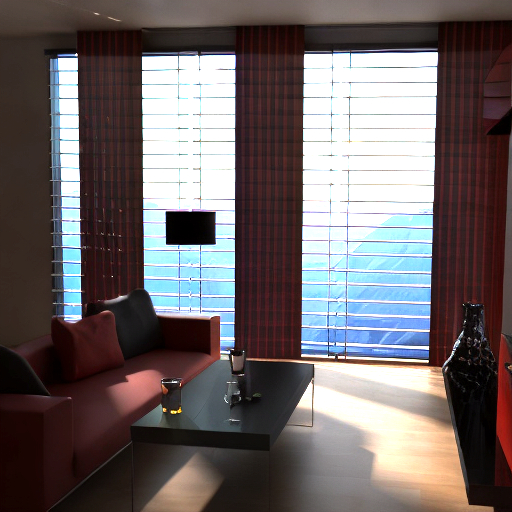}};
\begin{scope}[x={(img10.south east)}, y={(img10.north west)}]
\draw[gray!30,, thick] (0.42,0.13) rectangle (0.49,0.28);
\end{scope}
\end{tikzpicture}
&
\begin{tikzpicture}
\node[anchor=south west,inner sep=0] (img11) at (0,0)
{\includegraphics[width=0.16\textwidth]{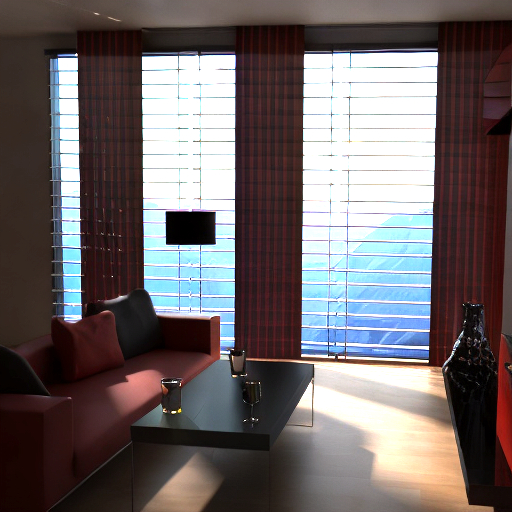}};
\begin{scope}[x={(img11.south east)}, y={(img11.north west)}]
\draw[gray!30,, thick] (0.45,0.13) rectangle (0.52,0.28);
\end{scope}
\end{tikzpicture}
&
\begin{tikzpicture}
\node[anchor=south west,inner sep=0] (img12) at (0,0)
{\includegraphics[width=0.16\textwidth]{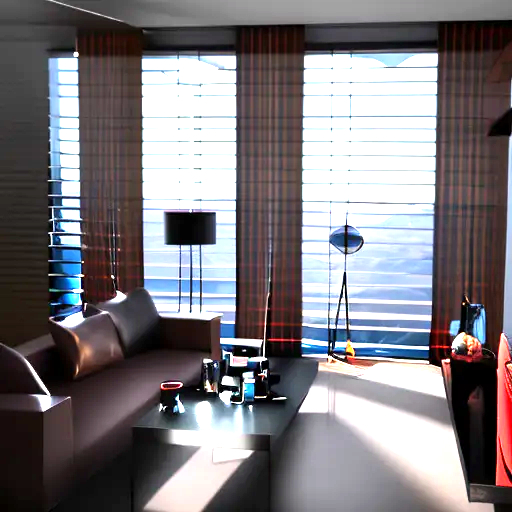}};
\begin{scope}[x={(img12.south east)}, y={(img12.north west)}]
\draw[gray!30,, thick] (0.45,0.13) rectangle (0.52,0.28);
\end{scope}
\end{tikzpicture}
\\
\begin{tikzpicture}
\node[anchor=south west,inner sep=0] (img13) at (0,0)
{\includegraphics[width=0.16\textwidth]{image/comparehandels/rendered_004.jpg}};
\begin{scope}[x={(img13.south east)}, y={(img13.north west)}]
\draw[gray!30,, thick] (0.25,0.1) rectangle (0.5,0.5);
\end{scope}
\end{tikzpicture}
&
\begin{tikzpicture}
\node[anchor=south west,inner sep=0] (img14) at (0,0)
{\includegraphics[width=0.16\textwidth]{image/comparehandels/frame_038.jpg}};
\begin{scope}[x={(img14.south east)}, y={(img14.north west)}]
\draw[gray!30,, thick] (0.43,0.2) rectangle (0.68,0.6);
\end{scope}
\end{tikzpicture}
&
\begin{tikzpicture}
\node[anchor=south west,inner sep=0] (img15) at (0,0)
{\includegraphics[width=0.16\textwidth]{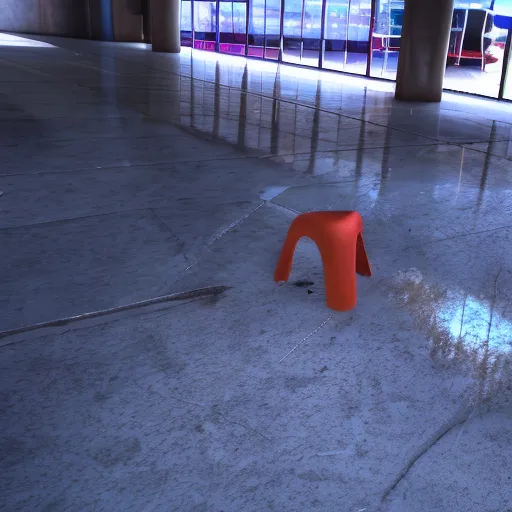}};
\begin{scope}[x={(img15.south east)}, y={(img15.north west)}]
\draw[gray!30,, thick] (0.48,0.2) rectangle (0.73,0.6);
\end{scope}
\end{tikzpicture}
\\
\end{tabular}
\caption{Comparison with Diffusion Handle\cite{pandey2024diffusion}. In this figure, we compare object movement results between our method and Diffusion Handle. Our approach consistently achieves higher-quality outputs with minimal background alterations, whereas Diffusion Handle exhibits more pronounced background changes and underperforms under extreme lighting conditions.}
\label{fig:comparisonmove}
\end{figure}

%% file: imagecompare2.tex
\begin{figure}[htbp]
\centering
\setlength{\tabcolsep}{-1.0pt} 
\renewcommand{\arraystretch}{1.0} 

\begin{tabular}{c c c} 

    \begin{tabular}{c}
        \textbf{Original} \\
        \begin{tikzpicture}
            \node[anchor=south west,inner sep=0] (img1) at (0,0)
                {\includegraphics[width=0.165\textwidth]{image/photo_inpaint_1.jpg}};
            \begin{scope}[x={(img1.south east)},y={(img1.north west)}]
            \end{scope}
        \end{tikzpicture}
    \end{tabular} &
    \begin{tabular}{c}
        \textbf{Ours} \\
        \begin{tikzpicture}
            \node[anchor=south west,inner sep=0] (img2) at (0,0)
                {\includegraphics[width=0.165\textwidth]{image/final_frame_model1_1.jpg}};
            \begin{scope}[x={(img2.south east)},y={(img2.north west)}]
                \draw[gray!30,,thick] (0.1,0.02) rectangle (0.8,0.4);
            \end{scope}
        \end{tikzpicture}
    \end{tabular} &
    \begin{tabular}{c}
        \textbf{RGBX} \\
        \begin{tikzpicture}
            \node[anchor=south west,inner sep=0] (img3) at (0,0)
                {\includegraphics[width=0.165\textwidth]{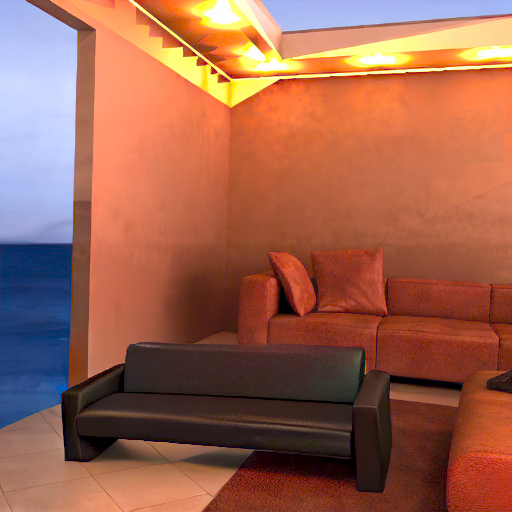}};
            \begin{scope}[x={(img3.south east)},y={(img3.north west)}]
                \draw[gray!30,,thick] (0.1,0.02) rectangle (0.8,0.4);
            \end{scope}
        \end{tikzpicture}
    \end{tabular} \\[10pt] 


    \begin{tabular}{c}
        \begin{tikzpicture}
            \node[anchor=south west,inner sep=0] (img7) at (0,0)
                {\includegraphics[width=0.165\textwidth]{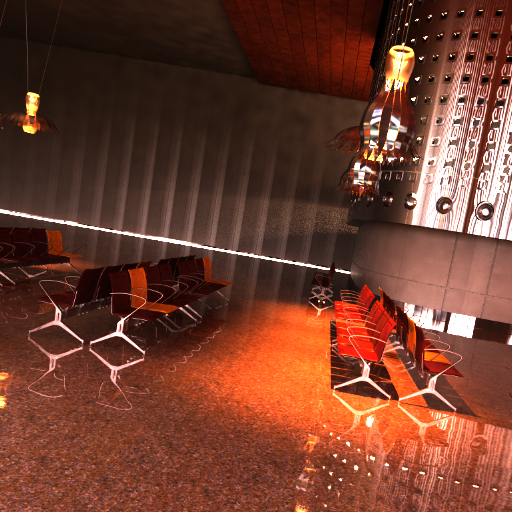}};
            \begin{scope}[x={(img7.south east)},y={(img7.north west)}]
            \end{scope}
        \end{tikzpicture}
    \end{tabular} &
    \begin{tabular}{c}
        \begin{tikzpicture}
            \node[anchor=south west,inner sep=0] (img8) at (0,0)
                {\includegraphics[width=0.165\textwidth]{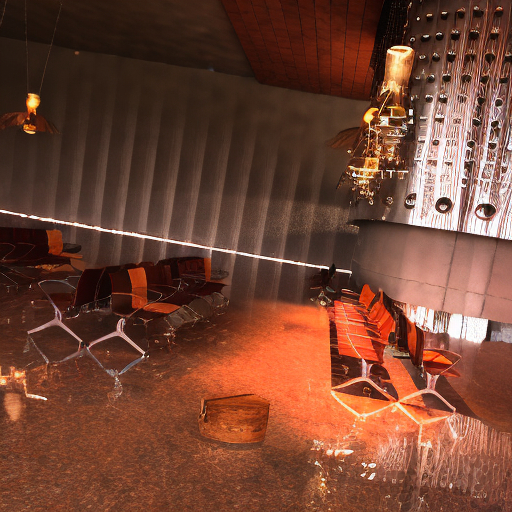}};
            \begin{scope}[x={(img8.south east)},y={(img8.north west)}]
                \draw[gray!30,,thick] (0.165,0.03) rectangle (0.6,0.3);
            \end{scope}
        \end{tikzpicture}
    \end{tabular} &
    \begin{tabular}{c}
        \begin{tikzpicture}
            \node[anchor=south west,inner sep=0] (img9) at (0,0)
                {\includegraphics[width=0.165\textwidth]{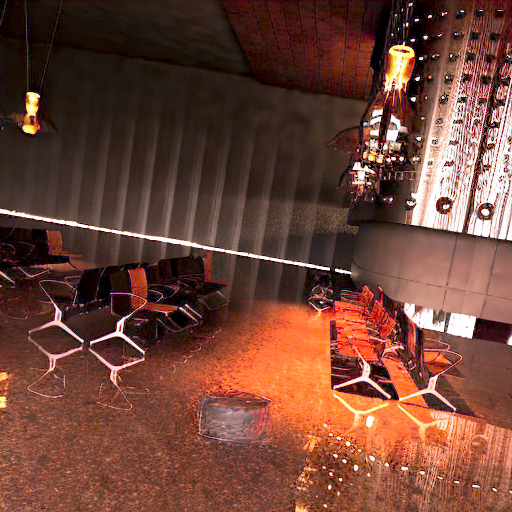}};
            \begin{scope}[x={(img9.south east)},y={(img9.north west)}]
                \draw[gray!30,,thick] (0.165,0.03) rectangle (0.6,0.3);
            \end{scope}
        \end{tikzpicture}
    \end{tabular} \\[10pt]
    

    \begin{tabular}{c}
        \begin{tikzpicture}
            \node[anchor=south west,inner sep=0] (img13) at (0,0)
                {\includegraphics[width=0.165\textwidth]{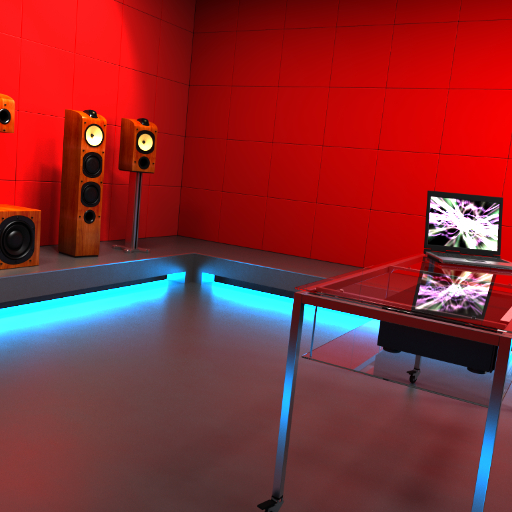}};
            \begin{scope}[x={(img13.south east)},y={(img13.north west)}]
            \end{scope}
        \end{tikzpicture}
    \end{tabular} &
    \begin{tabular}{c}
        \begin{tikzpicture}
            \node[anchor=south west,inner sep=0] (img14) at (0,0)
                {\includegraphics[width=0.165\textwidth]{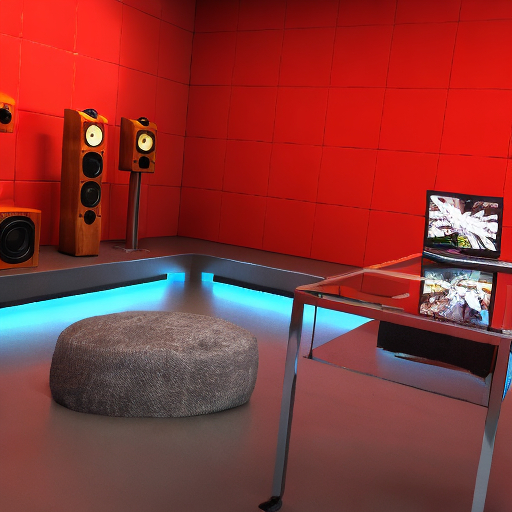}};
            \begin{scope}[x={(img14.south east)},y={(img14.north west)}]
                \draw[gray!30,,thick] (0.02,0.02) rectangle (0.55,0.5);
            \end{scope}
        \end{tikzpicture}
    \end{tabular} &
    \begin{tabular}{c}
        \begin{tikzpicture}
            \node[anchor=south west,inner sep=0] (img15) at (0,0)
                {\includegraphics[width=0.165\textwidth]{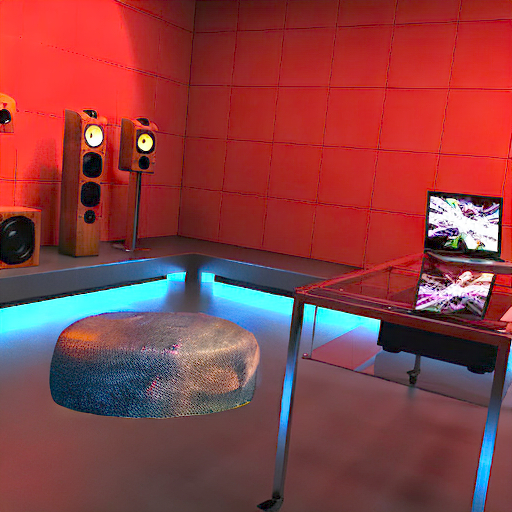}};
            \begin{scope}[x={(img15.south east)},y={(img15.north west)}]
                \draw[gray!30,,thick] (0.05,0.02) rectangle (0.55,0.5);
            \end{scope}
        \end{tikzpicture}
    \end{tabular} \\[10pt]

    \begin{tabular}{c}
        \begin{tikzpicture}
            \node[anchor=south west,inner sep=0] (img16) at (0,0)
                {\includegraphics[width=0.165\textwidth]{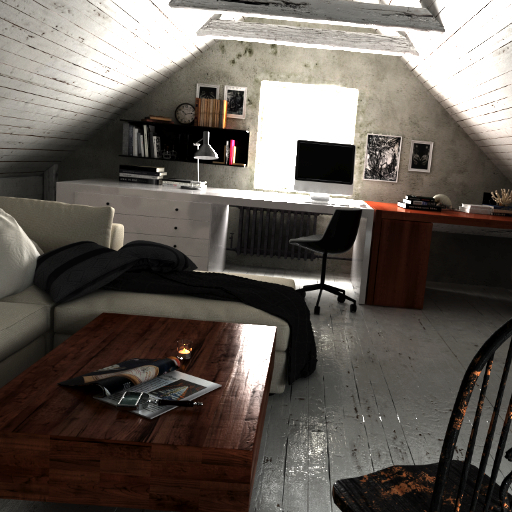}};
            \begin{scope}[x={(img16.south east)},y={(img16.north west)}]
            \end{scope}
        \end{tikzpicture}
    \end{tabular} &
    \begin{tabular}{c}
        \begin{tikzpicture}
            \node[anchor=south west,inner sep=0] (img17) at (0,0)
                {\includegraphics[width=0.165\textwidth]{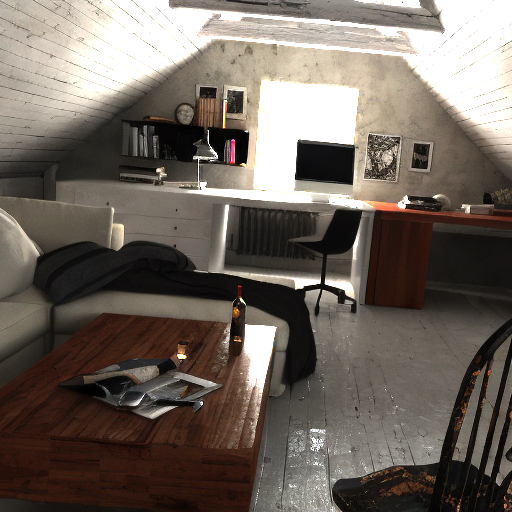}};
            \begin{scope}[x={(img17.south east)},y={(img17.north west)}]
                \draw[gray!30,,thick] (0.41,0.26) rectangle (0.52,0.48);
            \end{scope}
        \end{tikzpicture}
    \end{tabular} &
    \begin{tabular}{c}
        \begin{tikzpicture}
            \node[anchor=south west,inner sep=0] (img18) at (0,0)
                {\includegraphics[width=0.165\textwidth]{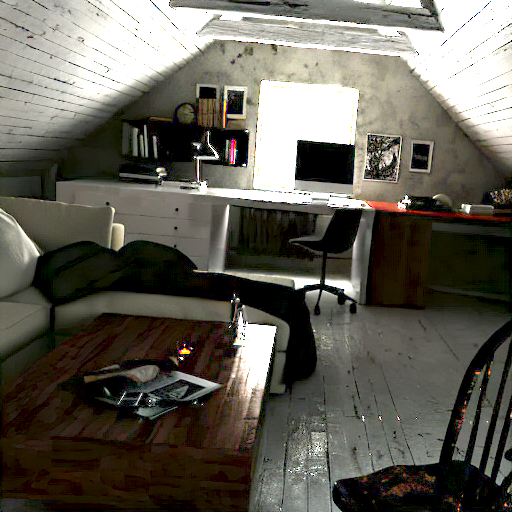}};
            \begin{scope}[x={(img18.south east)},y={(img18.north west)}]
                \draw[gray!30,,thick] (0.41,0.26) rectangle (0.52,0.48);
            \end{scope}
        \end{tikzpicture}
    \end{tabular} \\

\end{tabular}

\caption{Comparison with \revtwo{RGB$\leftrightarrow$X}~\cite{zeng2024rgb}. This figure presents an inpainting comparison between our method and an \revtwo{RGB$\leftrightarrow$X}-inpainting variant. The original images and inserted objects are synthetic data from the Hypersim dataset. Our approach demonstrates higher shadow quality and overall image fidelity compared to \revtwo{RGB$\leftrightarrow$X}.}
\label{fig:comparisonwithrgbx}
\end{figure}